\documentclass[12pt]{article}
\usepackage{ulem}
\usepackage{array}
\usepackage{multirow}
\usepackage{cite}
\usepackage{amsthm}
\usepackage{amsmath}
\usepackage{bigints}
\usepackage{amssymb}
\usepackage{tikz}
\usepackage{cite}
\usepackage{latexsym}
\usepackage{epsfig,graphicx,epstopdf}
\usepackage{tikz}
\usetikzlibrary{patterns}
\usepackage{float}
\usepackage{graphicx}
\usepackage{subcaption}
\usepackage{cite}
\usepackage{amsthm}
\usepackage{amsmath}
\usepackage{bigints}
\usepackage{amssymb}
\usepackage{tikz}
\usepackage{float}
\usepackage{graphicx}
\usetikzlibrary{patterns}
\usepackage{cite}
\usepackage{amsthm}
\usepackage{amsmath}
\usepackage{bigints}
\usepackage{amssymb}
\usepackage{graphicx,color}
\newcommand\ddfrac[2]{\frac{\displaystyle #1}{\displaystyle #2}}
\newcommand{\be}{\begin{equation}}

\newcommand{\bea}{\begin{eqnarray}}
\newcommand{\eea}{\end{eqnarray}}
\newcommand{\nn}{\nonumber}
\renewcommand{\(}{\left(}

\newcommand{\bc}{\begin{center}}
\newcommand{\ec}{\end{center}}
\newcommand{\p}{\partial}

\newcommand{\ds}{\displaystyle}
\newcommand{\beq}{\begin{eqnarray}}
\newcommand{\eeq}{\end{eqnarray}}
\newcommand{\beqq}{\begin{eqnarray*}}
\newcommand{\eeqq}{\end{eqnarray*}}

\newcommand{\eps}{\varepsilon}

\newcommand{\x}{\mbox{\boldmath$x$}}

\newcommand{\s}{\mbox{\boldmath$s$}}

\begin{document}
%%%%%%%%%%%%%%%%%%%%%%%%%%%%%%%%%%%%%%%%%%55
%\title{Heterogeneous cross-linked polymers of chromatin organization during cell differentiation}
\title{Voltage distribution in a non-locally but globally electroneutral confined electrolyte medium: applications for nanophysiology}
\author{A. Tricot$^{1}$ I.M. Sokolov$^2*$,  D. Holcman$^{1}$ \footnote{$^1$ Data Modeling, Computational Biology and Predictive Medicine, Ecole Normale Sup\'erieure, 46 rue d'Ulm 75005 Paris, France.$^2$ Institute of Physics and IRIS Adlershof, Humboldt University Berlin, Newtonstr. 15, 12489 Berlin, Germany.}}
\maketitle
%%%%%%%%%%%%%%%%%%%%%%%%%%%%%%%%%%%%%%%%%%
\begin{abstract}
The distribution of voltage in sub-micron cellular domains remains poorly understood.  In neurons, the voltage results from the difference in ionic concentrations which are continuously maintained by pumps and exchangers. However, it not clear how electro-neutrality could be maintained by an excess of fast moving positive ions that should be counter balanced by slow diffusing negatively charged proteins. Using the theory of electro-diffusion, we study here the voltage distribution in a generic domain, which consists of two concentric disks (resp. ball) in two (resp. three) dimensions, where a negative charge is fixed in the inner domain. When global but not local electro-neutrality is maintained, we solve the Poisson-Nernst-Planck equation both analytically and numerically in dimension 1 (flat) and 2 (cylindrical) and found that the voltage changes considerably on a spatial scale which is much larger than the Debye screening length, which assumes electro-neutrality. The present result suggests that long-range voltage drop changes are expected in neuronal microcompartments, probably relevant to explain the activation of far away voltage-gated channels located on the surface.
\end{abstract}
%%%%%%%%%%%%%%%%%%%%%%%%%%%%%%%%%%%%%%%%%%
\maketitle
%%%%%%%%%%%%%%%%%%%%%%%%%%%%%%%%%%%%%%%%%%
\section{Introduction}
%%%%%%%%%%%%%%%%%%%%%%%%%%%%%%%%%%%%%%%%%%55
How voltage and ionic concentrations are distributed and regulated in excitable cells such as neurons, astrocytes, etc.. remains a challenging question, despite decades of experimental and theoretical efforts \cite{Hille,eisenberg1970three,Bezanilla,Rall,YusteBook,HY2015,Sylantyev}. In particular, the voltage in microdomains such as initial segments, dendrites, dendritic spines, remain difficult to study experimentally due to their small size. The ionic concentrations are constantly regulated in order to maintain the physiological gradients: while potassium ions are extruded, sodium ions must be pumped in through energy dependent exchangers \cite{Hille}.  In recent years, the voltage distribution and the ionic currents have been measured using nanopipettes \cite{jayant2019flexible,jayant2017targeted} and voltage dyes \cite{cartailler2018deconvolution}. Neuronal microdomains are characterized by an excess of positive ions (sodium and potassium), not compensated by chloride. However the missing negative charges should be carried by heavy proteins and molecules inside the cytoplasm characterized by small diffusion coefficients compared to the ones of the main ions. Yet, the overall cytoplasmic medium is expected to be electroneutral, although measurements should be performed \cite{cartailler2018deconvolution,HY2015} in cellular domains such as dendritic spines, pre-synaptic terminal or glial protrusions.\\
The classical framework to study electrical properties of cytoplasm which are electrolytes is the electro-diffusion theory \cite{Hille,gillespie2002coupling,schuss2001derivation} which consists of modeling the motion of diffusing ions in water, where the electrostatic force is due to the charge concentration differences between positive and negative species.\\
In the classical Debye theory \cite{Hille}, the voltage of a charge immersed is estimating in a neutral electrolyte. This theory predicts a screening of an excess charge, due to the exponential decay of the electrical field. The theory is based on two main assumptions 1) the field induced by the excess charge is small compared to thermal fluctuations and 2) a strict electroneutrality condition imposed at infinity, where the concentration of positive and the negative charges are equal far away of the immersion of the test volume. The Debye characteristic length is $\lambda_D=\left( \frac{\varepsilon \varepsilon_0 k_B T}{z^2 e^2 N_A c_0} \right)^{\frac12}$, for the electron charge $e$, The temperature $T$, the Boltzmann constant $k_B$, the valence $z$, the vacuum permittivity  $\varepsilon_0$ and $\varepsilon$ the relative permittivity of the ions, the avogadro number $N_A$  and the concentration of ions $c_0$ . \\
In the extreme case of non-electrical medium, theoretical analysis and numerical simulations revealed a long-range log-decay of the electric field \cite{berg2011analytical,berg2009exact,PhysD2016,NonLin2017} and a modulation of the voltage distribution due to an oscillating \cite{cartailler2017electrostatics} or a cusp \cite{NonLin2017,cartailler2019steady} geometry. \\
In this manuscript, we compute the voltage and charge distribution when the condition of global but not local electro-neutrality is maintained. We consider a ball containing positive and negative charges, however a fraction of negative charges is fixed in the inner ball (Fig. \ref{fig1}). The external boundary does not allow charges to escape.  The manuscript is organized as follows: in section 1, we present general PNP model. We summarize our main results in table 1. In section 2, we treat the case of one dimension. We solve the PNP equation using elliptic integrals and obtain the decay of the voltage near the boundary. In section 3, we study the solutions in dimensions two and three.  We determine the voltage and charge distribution when we vary the static negative charges.
%%%%%%%%%%%%%%%%%%%%%%%%%%%%%%%%%%%%%%%%%%%%%%%%%%%%%%
\section{Model of global but not local electroneutrality}\label{s:2}
\begin{figure}[http!]
  \centering
  \begin{tikzpicture}
    \draw[blue, very thick] (0,0) circle (80pt);
    \draw[pattern = north west lines,lightgray] (0,0) circle (20pt);
    \filldraw[gray, thick] (0,0) circle (8pt);
    \draw[red, very thick] (0,0) circle (20pt);
    \draw[->] (0,0) -- (0,2.75) node[midway, right] {$R$};
    \draw[->] (0,0) -- (0.65,0) node[midway, above] {$R_0$};
    \node (Q) at (0,0) [label = below:$Q^-$]{};
    \node (rom) at (-0.9, -1.7) {\bf $q^-, \rho^-, n^-$};
    \node (rop) at (-1.7, -0.5) {\bf $q^+, \rho^+, n^+$};
    \node (omega) at (2.3,2.3) {\bf $\Omega$};
    \node (omega) at (0.7,0.7) {\bf $\Omega_0$};
\end{tikzpicture}
\caption{Schematic representation of the geometry $\Omega$ made of two concentric disks: the small one $\Omega_0$ containing the fixed charged $Q^- = -N z e$, modeling impenetrable proteins (red circle). Between the red boundary and the blue one, negative charges (total charge $q^- = -n^- ze$, that could represent chloride ions) is mixed in water with positive ions  $q^+ = n^+ ze$, representing potassium and sodium. The global electro-neutrality imposes $Q^- + q^- + q^+ = 0$ (equivalently $N + n^- = n^+$).}
  \label{fig1}
\end{figure}
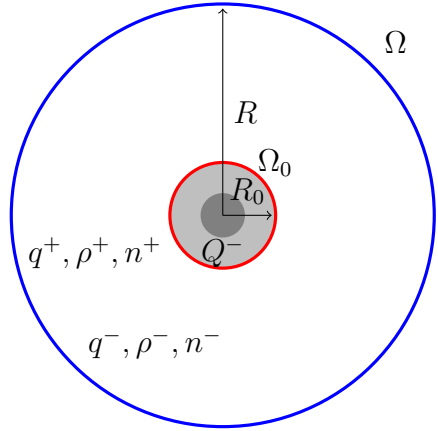

%%%%%%%%%%%%%%%%%%%%%%%%%%%%%%%%%%%%%%%%%%%%%%%%%%%%%%%%%%
To model global but not local electro-neutrality, we use an elementary geometry of a domain $\Omega$ consisting in two concentric disks in dimension two and balls in dimension three. We impose a negative charge inside (Fig. \ref{fig1}).
%\begin{tikzpicture}
%
%\draw[blue, very thick] (0,0) rectangle (3,2);
%\draw[orange, ultra thick] (4,0) -- (6,0) -- (5.7,2) -- cycle;
%
%\end{tikzpicture}
%
%
%\begin{tikzpicture}
%\draw[gray, thick] (-1,2) -- (2,-4);
%\draw[gray, thick] (-1,-1) -- (2,2);
%\filldraw[black] (0,0) circle (2pt) node[anchor=west] {Intersection point};
%\end{tikzpicture}
%%%%%%%%%%%%%%%%%%%%%%%%%%%%%%%%%%%%%%%%%%%%%%%%%%%%%%
%%%%%%%%%%%%%%%%%%%%%%%%%%%%%%%%%%%%%%%%%%%%%%%%%%%%%%
%%%%%%%%%%%%%%%%%%%%%%%%%%%%%%%%%%%%%%%%%%%%%%%%%%%%%%
\subsection{The Poisson-Nernst-Planck equations in the domain $\Omega$}\label{ss:2}
%%%%%%%%%%%%%%%%%%%%%%%%%%%%%%%%%%%%%%%%%%%%%%%%%%%%%%
The coarse-grain Poisson-Nernst-Planck equations model electro-diffusion \cite{Hille,Singer,schuss2001derivation} in a electrolyte. In the domain $\Omega$ (Fig. \ref{fig1}), the total charge is the sum of mobile positive $n^+$ and negative $n^-$ charges plus a fixed number negative charges $N^{static}=N$ located in an impenetrable (unaccessible) subregion $\Omega_0 \subset  \Omega$ (red circle in Fig. \ref{fig1}), representing  negatively charged proteins.  We assume global electroneutrality:
\beq \label{globalRelation}
N + n^- = n^+.
\eeq
For a ionic valence $z$, the total number of particles is
\beq
\int \limits_{\Omega-\Omega_0} \rho_p(\tilde\x,t) d\tilde\x=n^+, \, \int \limits_{\Omega-\Omega_0} \rho_n(\tilde\x,t) d\tilde\x=n^-
\eeq
and thus the total charge is
\beqq
q_{\pm}=\pm zeN,\,  Q^- + q^- + q^+ = 0
\eeqq
where $e$ is the electron charge. The particle density $\rho_p(\tilde\x,t)$ is the solution of the Nernst-Planck equation
\beq
D\left[\Delta \rho_p(\tilde\x,t) +\frac{ze}{kT} \nabla\cdot \left(\rho_p(\tilde\x,t) \nabla \phi(\tilde\x,t)\right)\right]&=&\,
\frac{\p\rho_p(\tilde\x,t)}{\p t}\hspace{0.5em}\mbox{for}\ \tilde\x\in \Omega-\Omega_0\label{NPE} \nonumber \\
D\left[\Delta \rho_n(\tilde\x,t) -\frac{ze}{kT} \nabla\cdot \left(\rho_n(\tilde\x,t) \nabla \phi(\tilde\x,t)\right)\right]&=&\,
\frac{\p\rho_n(\tilde\x,t)}{\p t}\hspace{0.5em}\mbox{for}\ \tilde\x\in \Omega-\Omega_0\label{NPE2} \nonumber\\
D\left[\frac{\p\rho(\tilde\x,t)}{\p n}+\frac{ze}{kT}\rho(\tilde\x,t)\frac{\p\phi(\tilde\x,t)}{\p
	n}\right]&=&\,0\hspace{0.5em}\mbox{for}\ \tilde\x\in\p \Omega -\p \Omega_0 \label{noflux}\\
\rho(\tilde\x,0)&=&\, {\rho_0}(\tilde\x)\hspace{0.5em}\mbox{for}\ \tilde\x\in\tilde\Omega,\label{IC}
\eeq
where $kT$ represents the thermal energy.  The electric potential $\phi(\tilde\x,t)$ in $\tilde\Omega$ satisfies the Poisson equation
\begin{align}
\label{poisson} \Delta \phi(\tilde\x,t) =&\,
-\frac{ze\left(\rho_p(\tilde\x,t)-\rho_n(\tilde\x,t)\right)}{\eps_r\eps_0}\hspace{0.5em}\mbox{for}\ \tilde\x\in\tilde\Omega\\
\frac{\p\phi(\tilde\x,t)}{\p n}=&\,-\tilde\sigma(\tilde\x,t)\hspace{0.5em}\mbox{for}\ \tilde\x\in{\p \Omega-\Omega_0},\label{Boundary_Phi} % -\frac{\sigma(\x,t)}{\eps\eps_0}
\end{align}
where $\eps_r\eps_0$ is the permittivity of the medium and $\tilde\sigma(\tilde\x,t)$ is the surface charge density on the boundary $\p\Omega$.
%%%%%%%%%%%%%%%%%%%%%%%%%%%%%%%%%%%%%%%%%%%%%%%%%%%%%%
\subsection{Steady-state solution}\label{ss:SSS}
%%%%%%%%%%%%%%%%%%%%%%%%%%%%%%%%%%%%%%%%%%%%%%%%%%%%%%
To study the effect of non-local electroneutrality,  we study the solution of the steady-state equation \eqref{NPE} in the normalized domain $\tilde\Omega$ (of radius 1). The Boltzmann distributions are given by
\beq
\rho_p(\tilde\x)=n^+\frac{\exp\left\{-\ds\frac{ze\phi(\tilde\x)}{kT}\right\}}
{\int_{\tilde\Omega-\tilde\Omega_0}\exp\left\{-\ds\frac{ze\phi(\x)}{kT}\right\}\,d \x},\label{N} \\
\rho_n(\tilde\x)=n^-\frac{\exp\left\{\ds\frac{ze\phi(\tilde\x)}{kT}\right\}}
{\int_{\tilde\Omega-\tilde\Omega_0}\exp\left\{\ds\frac{ze\phi(\x)}{kT}\right\}\,d\x},\label{N2}
\eeq
hence \eqref{poisson} results in the nonlinear Poisson equation
\beq
\Delta\phi(\tilde\x)=-\frac{zeN_p\exp\left\{-\ds\frac{ze\phi(\tilde\x)}{kT}
\right\}}{\eps_r\eps_0{\ds\int_{\tilde\Omega-\tilde\Omega_0}	\exp\left\{-\ds\frac{ze\phi(\s)}{kT}\right\}\,d\s}}+ \frac{zeN_n\exp\left\{\ds\frac{ze\phi(\tilde\x)}{kT}\right\}}
{\eps_r\eps_0{\ds\int_{\tilde\Omega-\tilde\Omega_0}\exp\left\{\ds\frac{ze\phi(\s)}{kT}\right\}\,d\s}}.\label{Deltaphi}
\eeq
In region $\tilde\Omega_1$, the Poisson equation is
\beq
\Delta\phi(\tilde\x)= -\frac{ze N}{\eps_r\eps_0 V_1},
\eeq
and thus
\beq\label{compatibility}
\int_{\Sigma_0}\frac{\p\phi({\tilde\x})}{\p n} dS_{\x}= -\frac{ze N }{\eps_r\eps_0},
\eeq
where $\Sigma_0$ is the boundary of $\Omega_0$. The global electro-neutrality (relation \ref{globalRelation}) leads to the compatibility condition imposed by Gauss flux integral
\beq\label{compatibility2}
\int_{\Sigma_0}\frac{\p\phi({\tilde\x})}{\p n} dS_{\x}-\int_{\Sigma}\frac{\p\phi({\tilde\x})}{\p n} dS_{\x}= n^+ -n^-.
\eeq
Thus,
\beq
\int_{\Sigma}\frac{\p\phi({\tilde\x})}{\p n} dS_{\x}=0.
\eeq
By symmetry, we impose that  $\ds \frac{\p \phi}{\p n}$ is constant on the two surfaces $\Sigma$ and $\Sigma_0$ and thus we impose the conditions:
\beq\label{compatibility3}
\frac{\p\phi({\tilde\x})}{\p n}&=&-\frac{ze N }{\eps_r\eps_0 |\Sigma_1|}\,\mbox{  for  }\, \tilde\x\in \Sigma_0. \\
\frac{\p\phi({\tilde\x})}{\p n}&=&0\,\mbox{  for  }\, \tilde\x\in \Sigma.
\eeq
In spherical symmetry, the Poisson's equation \eqref{Deltaphi} reduces to
\beq\label{eq:pnp1}
\Phi''(r) + \frac{d-1}{r} \Phi'(r) =
\frac{ze}{\varepsilon \varepsilon_0 S_d}
&\left(
n^-\ddfrac{\exp \left( \ddfrac{ze\Phi(r)}{k_B T} \right)}{\int_{R_0}^{R} \exp \left( \ddfrac{ze\Phi(r)}{k_B T} \right) r^{d-1} dr} \right.
\\
&- \left. n^+\ddfrac{\exp \left( -\ddfrac{ze\Phi(r)}{k_B T} \right) }{\int_{R_0}^{R} \exp \left( -\ddfrac{ze\Phi(r)}{k_B T} \right) r^{d-1} dr}
\right).
\eeq
We normalize the radius by setting $r = Rx$ for $a \leq x \leq 1$ where $a = \frac{R_0}{R}$. Here
\beq
 u = \frac{ze}{k_B T} \Phi,
\quad \lambda_d = \frac{(ze)^2}{S_d R^{d-2} \varepsilon \varepsilon_0 k_B T}
\eeq
Here $S_d$ is the surface area of the unit sphere in $\mathbb{R}^d$. Eq.\eqref{eq:pnp1} becomes
\beq \label{eq:pnpgeneral}
 u''(x) + \frac{d-1}{x} u'(x) = I_\lambda e^{u(x)} - J_\lambda e^{-u(x)},
\eeq
where we use the notations
\beq \label{eq:defIJ}
I_\lambda = \ddfrac{n^- \lambda_d}{\int_{a}^{1} \exp \left( \ddfrac{ze\Phi(x)}{k_B T} \right) x^{d-1} dx},\quad
J_\lambda = \ddfrac{n^+ \lambda_d}{\int_{a}^{1} \exp \left( - \ddfrac{ze\Phi(x)}{k_B T} \right) x^{d-1} dx}.
\eeq
We shall study the anionic $I_\lambda$ and cationic $J_\lambda$ strengths vs $\lambda$ and the solution $u$.
 %Using non-dimensional variables, we define the normalized field
%\beq \label{conversion}
%\bar u(\tilde\x)=\ds \frac{ze \phi({\tilde\x})}{kT},\quad\lambda= \frac{(ze)^2N}{\eps_r\eps_0 kT},
%\eeq
%where $\lambda$ generalizes the Bjerrum length $l_B=e^2/kT$.
Our goal here is to determine $u$ over the ball $\tilde \Omega$. The condition $u'(1) = 0$ is satisfied due to the global electro-neutrality. We impose that the voltage is zero on $\Sigma$, as it is defined to an additive constant. In summary the boundary conditions are
\beq \label{eq:bc}
u(1) = 0, \quad u'(1) = 0.
\eeq
Eq.\eqref{eq:pnpgeneral} and the boundary conditions in eq.\eqref{eq:bc} together form a one dimensional boundary value problem with the following properties:
\begin{itemize}
\item the derivative $u'$ is maximal at point $a$ and decreases toward $u'(1) = 0$.
\item $u$ is minimal at $x=a$ and increases toward $u(1) = 0$.
\item $u''(1) = I_\lambda - J_\lambda \leq 0$ i.e. $J_\lambda \geq I_\lambda$.
\end{itemize}
The strategy to find the solution is the following: since the parameters $I_\lambda$ and $J_\lambda$ depend on the solution $u$, we will first search for an analytical solution for any value of the parameter $\lambda$. We will then self-consistently compute the expression of $I_\lambda$ and $J_\lambda$. This steps imposes some restriction and we will show that solutions exist only for specific values of $(I_\lambda, J_\lambda)$.
%%%%%%%%%%%%%%%%%%%%%%%%%%%%%%%%%%%%%%%%%
\section{steady-state Solution of PNP eq.\eqref{eq:pnpgeneral} in flat geometry (dimension 1)}
%%%%%%%%%%%%%%%%%%%%%%%%%%%%%%%%%%%%%%%%%%
%%%%%%%%%%%%%%%%%%%%%%%%%%%%%%%%%%%%%%%%%
\begin{figure}[http!]
\includegraphics[width=\textwidth]{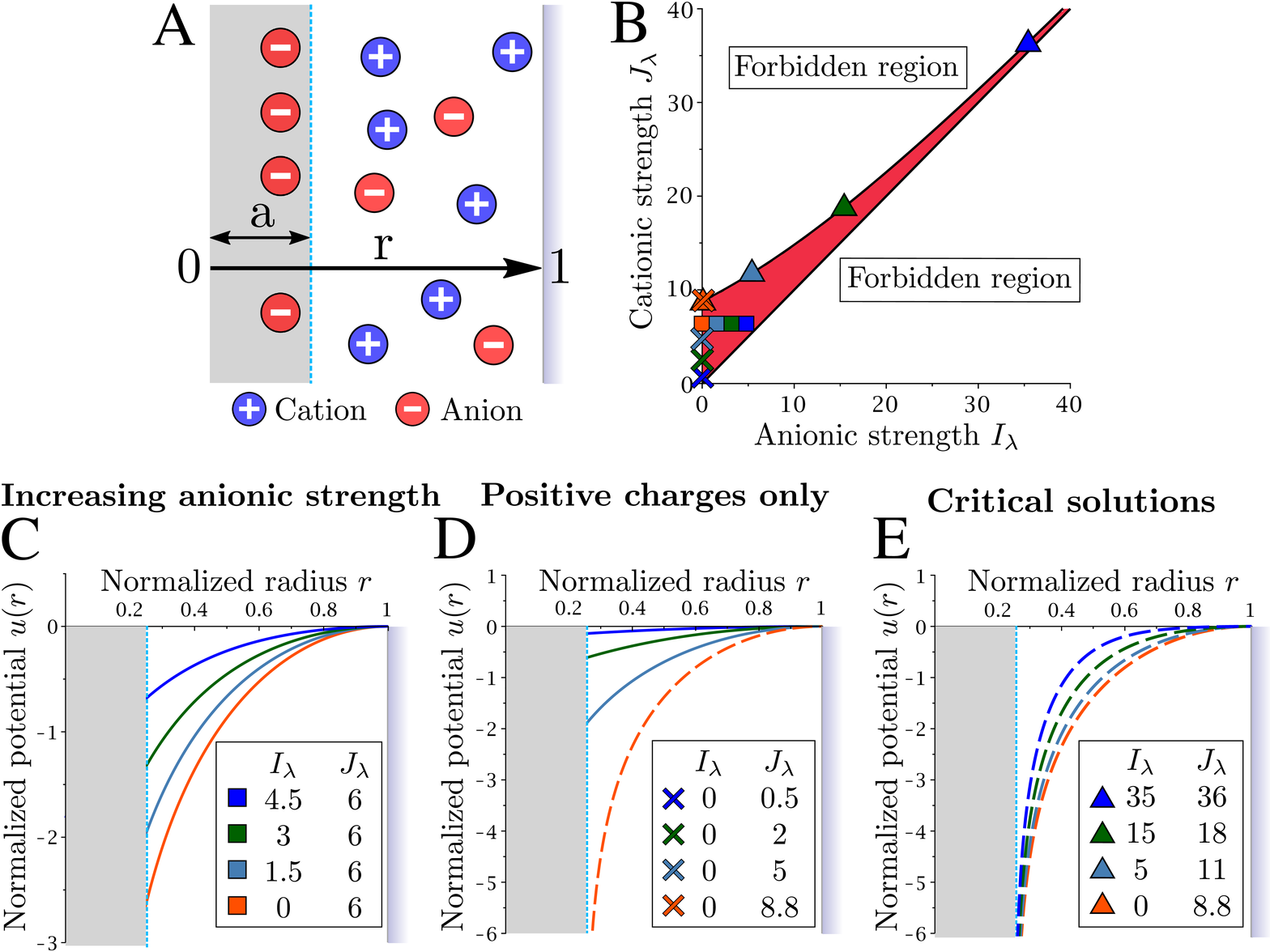}
\caption{Normalized potential $u(r)$ in dimension 1 for $\tilde{\Omega}=[0,1]$. \textbf{A.} Schematic representation of the domain.
\textbf{B.} Allowed (red) and forbidden regions for the parameters $I_\lambda$ and $J_\lambda$. The squares, crosses and triangles respectively refer to the curves on panels \textbf{C}, \textbf{D} and \textbf{E}.
\textbf{C.} Solution $u(r)$ with constant cationic strength $J_\lambda$ and increasing anionic strength $I_\lambda$. By increasing $I_\lambda$, the amplitude of the solution decays. \textbf{D.} No negative ions in the region $[0,a]$. The critical solution (dashed) develops a singularity in $r=a$.\textbf{E.} Critical solutions with a singularity at $r=a$. The parameters $I_\lambda$ and $J_\lambda$ are on the boundary of the red domain satifying relation $\sqrt{I_\lambda} + \sqrt{J_\lambda} = \frac{\sqrt{2} K(k)}{1-a}$.}
\label{fig:pnp_dim1}
\end{figure}
\begin{figure}[http!]
\includegraphics[scale = 0.3]{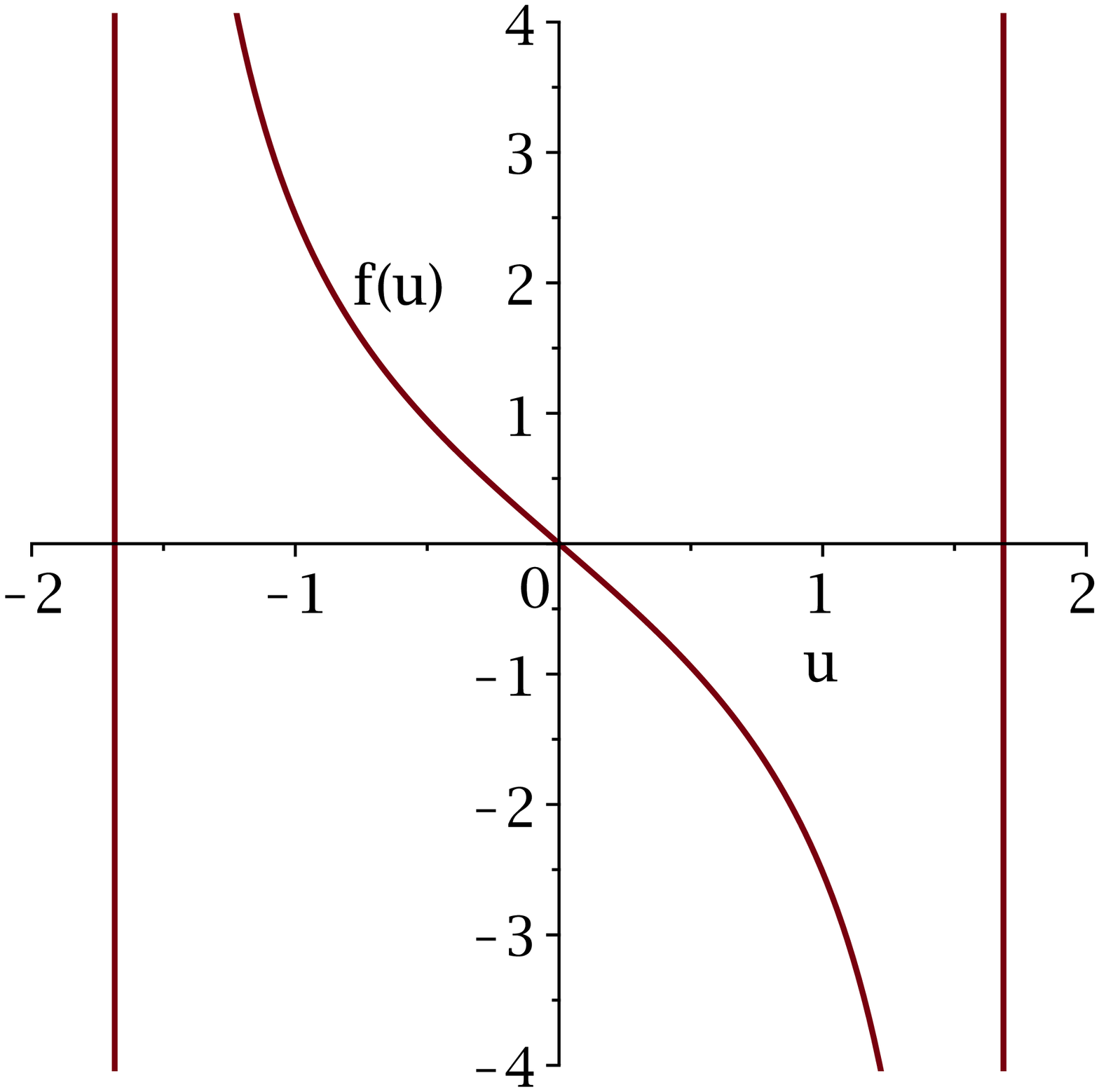}
\includegraphics[scale = 0.3]{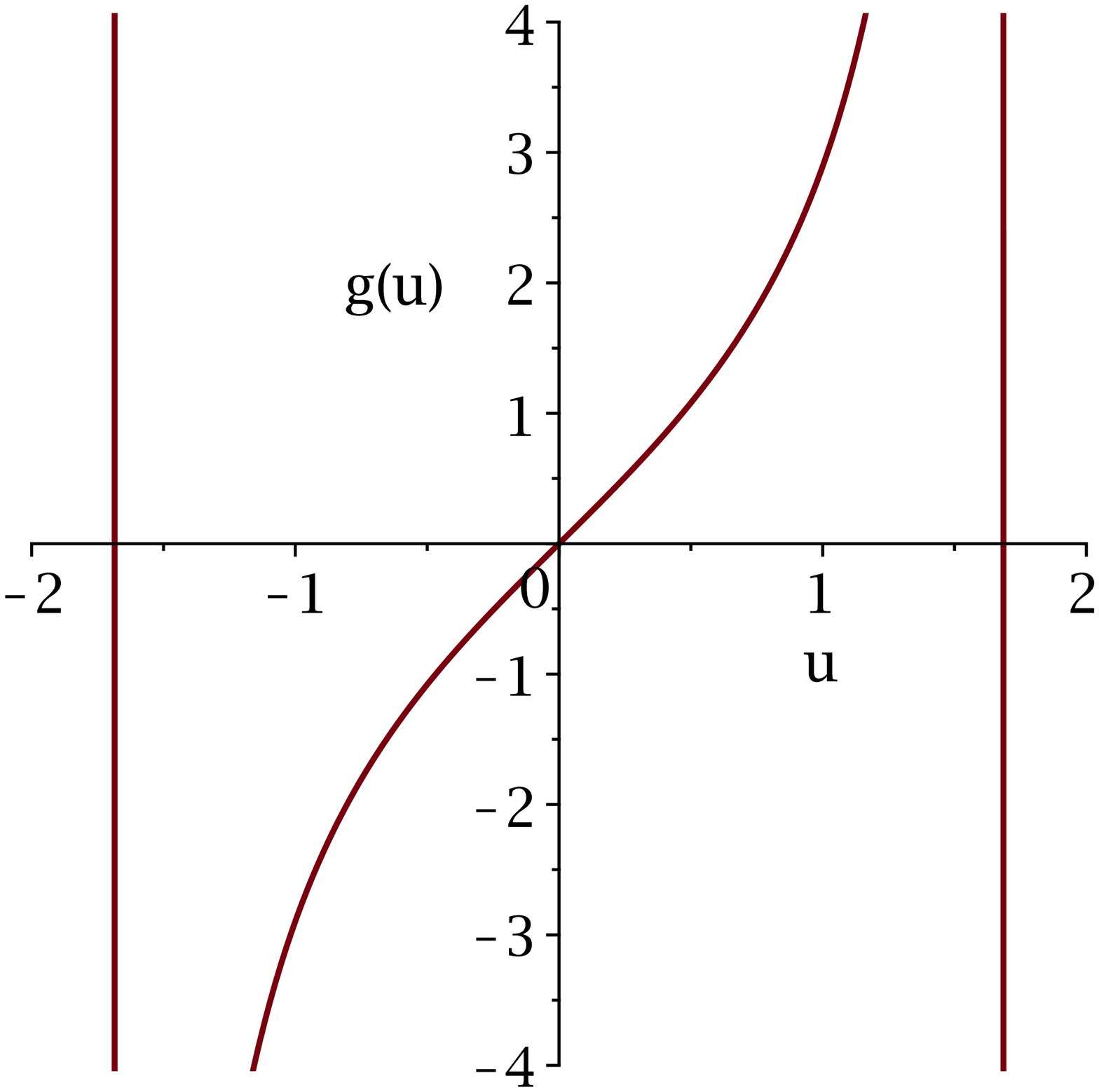}
\caption{Graph of the function $f$ and $g$ for $k = 1/2$. The asymptote is located for $u = K(k)$ (here $K\left( \frac12 \right) \approx 1.68$).}
\label{fig:graph_fg}
\end{figure}

%%%%%%%%%%%%%%%%%%%%%%%%%%%%%%%%%%%%%%%%%
In dimension 1, the normalized domain $\tilde{\Omega}$ is the interval $\tilde{\Omega}=[0,1]$ (Fig. \ref{fig:pnp_dim1}A). The fixed negative charges are located in $[0,a]$  while the mobile ions are in $a < x < 1$. The boundary value problem eq.\eqref{eq:pnpgeneral} reduces to
\beq\label{eq:bc_dim1}
u''(x) &=& I_\lambda e^{u(x)} - J_\lambda e^{-u(x)} \qquad \text{for $a < x < 1$}, \\
u(1) &=& 0, \quad u'(1) = 0. \nn
\eeq
We show by direct integration in Appendices \ref{appendix1} and \ref{appendix2} that the general solution can be expressed in terms of the Jacobian elliptic functions \cite{Abramowitz}
{\small
\beq\label{eq:dim1_u}
u(x)=-2 \ln\left( \frac12 \frac{\sqrt{I_\lambda} + \sqrt{J_\lambda}}{\sqrt{J_\lambda}}
\left(\operatorname{dc}\left(\frac{\sqrt{I_\lambda} + \sqrt{J_\lambda}}{\sqrt{2}} (x-1) \right)
 + \sqrt{1 - k_\lambda^2} \operatorname{nc}\left(\frac{\sqrt{I_\lambda} + \sqrt{J_\lambda}}{\sqrt{2}} (x-1) \right)
 \right) \right),
\eeq
}
where $\operatorname{dc}$ and $\operatorname{nc}$ are the elliptic functions \cite{Abramowitz} of modulus
\beq\label{eq:k_first}
k_\lambda = \sqrt{\frac{2}{1-c}} = \ddfrac{2\sqrt[4]{I_\lambda J_\lambda}}{\sqrt{I_\lambda} + \sqrt{J_\lambda}},
\eeq
with  $0 < k \leq 1$. The parameters $I_\lambda$ and $J_\lambda$ satisfy the inequality (Appendix \ref{appendix1})
\beq \label{eq:limitIJ}
\sqrt{I_\lambda} + \sqrt{J_\lambda} \leq \frac{\sqrt{2} K(k)}{1-a},
\eeq
where $K(k)$ is the complete elliptic integral of the first kind (Appendix \ref{appendix1}). The possible region (red in Fig. \ref{fig:pnp_dim1}B) is obtained by combining  conditions \ref{eq:k_first} and \ref{eq:limitIJ}. We plotted the solutions for various positive and negative charges (Fig. \ref{fig:pnp_dim1}C-E). In the boundary of validity (eq.\eqref{eq:limitIJ}), the solution $u$ develops a log-singularity at $x=a$ (Fig. \ref{fig:pnp_dim1} D-E, dashed lines). This situation is similar to the case of a single charge in the entire ball \cite{PhysD2016,NonLin2017}. It is interesting to observe the long -range voltage changes in this non-local electro-neutral medium, even in the limit of $a$ small (size of the impenetrable region containing negative charges).
%%%%%%%%%%%%%%%%%%%%%%%%%%%%%%%%%%%%%%%%%%%%%
To obtain a closed form of the solution, we compute  $I_\lambda$ and $J_\lambda$ (relation \ref{eq:defIJ}) with respect to the parameters $\lambda_d n^+$, $\lambda_d n^-$ and $a$.  A direct integration of the function $e^{u(x)}$ and $e^{-u(x)}$ over the interval $[a,1]$ (In appendix \ref{appendix3}) gives with
\beq
u_a = \frac{\sqrt{I_\lambda} + \sqrt{J_\lambda}}{\sqrt{2}} (a-1)
\eeq
that
\beq%\label{eq:SIJ_appendix}
\ds \lambda_d n^- &= \frac{\sqrt{I_\lambda} + \sqrt{J_\lambda}}{2\sqrt{2}} \left( f_{k_{\lambda}}(u_a) + \frac{ \sqrt{J_\lambda} - \sqrt{I_\lambda} }{\sqrt{I_\lambda} + \sqrt{J_\lambda}} g(u_a) \right),\label{eq:SI_appendix1} \\
\ds \lambda_d n^+ &= \frac{\sqrt{I_\lambda} + \sqrt{J_\lambda}}{2\sqrt{2}} \left( f_{k_{\lambda}}(u_a) - \frac{ \sqrt{J_\lambda} - \sqrt{I_\lambda} }{\sqrt{I_\lambda} + \sqrt{J_\lambda}} g(u_a) \right),\label{eq:SJ_appendix1}
\eeq
where we defined the two functions (Fig. \ref{fig:graph_fg})
\beq \label{eq:fg}
f_{k}(x) &=& 2 \operatorname{E}(x) - (2-k^2)x -2 \operatorname{sn}(x)\operatorname{dc}(x),\\
g(x) &=& 2\operatorname{sc}(x), \hbox{ for } x \in ]-K(k);K(k)[.
\eeq
Note that we can write (from relation \ref{eq:k_first})
\beq \label{klambda}
1-k_\lambda^2 = (\frac{ \sqrt{J_\lambda} - \sqrt{I_\lambda} }{\sqrt{I_\lambda} + \sqrt{J_\lambda}})^2.
\eeq
The parameter $k_\lambda$ represents the balance between the negative charges. Indeed,
\begin{itemize}
\item $k_\lambda \longrightarrow 1$, $ I_\lambda \approx J_\lambda$ and eq.\eqref{eq:SN} implies \textbf{$N \longrightarrow 0$}.
\item $k_\lambda \longrightarrow 0$, $ J_\lambda \gg I_\lambda$. Using $I_\lambda$ in eq.\eqref{eq:defIJ}, we get  \textbf{$n^- \longrightarrow 0$}.
\end{itemize}
%%%%%%%%%%%%%%%%%%%%%%%%%%%%%%%%%%%%%%%%%%
%%%%%%%%%%%%%%%%%%%%%%%%%%%%%%%%%%%%%%%%%%
Finally, the global electro-neutrality condition leads to the relation
\beq\label{eq:SN}
\lambda_d N = - \frac{ \sqrt{J_\lambda} - \sqrt{I_\lambda} }{\sqrt{2}} g(u_a).
\eeq
%We can rewrite the system of equation as
%\beq
%\ds 2(a-1) (\lambda_d ( n^-+\frac{N}{2})= u_a f(u_a) \\
%\lambda_d N = - \frac{ \sqrt{J_\lambda} - \sqrt{I_\lambda} }{\sqrt{2}} g(u_a).
%\eeq
%The first equation leads to the value of $u_a$ and the second to the difference $\sqrt{J_\lambda} - \sqrt{I_\lambda}$.
To conclude, for each positive and negative density ($n^+, n^-$) satisfying electroneutrality \ref{globalRelation}, the system of equations \ref{eq:SI_appendix1}-\ref{eq:SJ_appendix1}-\ref{klambda}-\ref{eq:SN} can be resolved and there is a unique couple ($J_\lambda, I_\lambda$) for which condition \eqref{eq:limitIJ} is satisfied, and thus the solution $u(x)$ is defined on the entire interval $[a,1]$.
%%%%%%%%%%%%%%%%%%%%%%%%%%%%%%%%%%%%%%%%%%%%%%%%5
\subsection{Explicit expressions for the difference of potential $u(1) - u(a) $}
%%%%%%%%%%%%%%%%%%%%%%%%%%%%%%%%%%%%%%%%%%%%%%%%%
We study here the potential difference between the surfaces of the two balls.
\beq  \label{eq:potential_dim1}
u(1) - u(a) = 2 \ln \left( \frac12 \frac{\sqrt{I_\lambda} + \sqrt{J_\lambda}}{\sqrt{J_\lambda}} \left( \operatorname{dc}(u_a) + \sqrt{1-k_{\lambda}^2}\operatorname{nc}(u_a) \right)\right),
\eeq
where
\beq
u_a = \frac{\sqrt{I_\lambda} + \sqrt{J_\lambda}}{\sqrt{2}} (a-1).
\eeq
%%%%%%%%%%%%%%%%%%%%%%%%%%%%%%%%%%%%%%%%5
%\begin{figure}[http!]
%  \centering
%  \includegraphics[scale=0.45]{potential3d.eps}
%  \caption{Difference in potential between the surfaces of the two balls as a function of $I_\lambda$ and $J_\lambda$}
%  \label{fig:potential_diff}
%\end{figure}
%%%%%%%%%%%%%%%%%%%%%%%%%%%%%%%%%%%%%%%%%5
The potential difference $u(1) - u(a)$ has a minimum when $I_\lambda = J_\lambda$ and grows with the difference between $I_\lambda$ and $J_\lambda$. The limit value for this difference depends on the value of the sum $\sqrt{I_\lambda} + \sqrt{J_\lambda}$ as shown by eq.\eqref{eq:limitIJ}. We shall now study some limit cases for the potential difference  $u(1) - u(a)$.
%%%%%%%%%%%%%%%%%%%%%%%%%%%%%%%%%%%%%%%%5
\subsubsection{Case $n^- \longrightarrow 0$ ($n^+ = N$)} \label{n-0}
%%%%%%%%%%%%%%%%%%%%%%%%%%%%%%%%%%%%%%%%5
In the case $n^- = 0$, we have $I_\lambda = 0$ (eq.\eqref{eq:defIJ}) and $k_{\lambda} = 0$. From eq.\eqref{eq:SJ_appendix1}, we obtain
\beq \label{eq:k0_eq1}
\lambda_d n^+ = \frac{\sqrt{J_\lambda}}{2\sqrt{2}} \left( f(u_a) - g(u_a) \right).
\eeq
When $k=0$, the Jacobian elliptic functions simplifies to trigonometric functions
\beq
f(u_a) = -2 \tan(u_a), g(u_a) = 2 \tan(u_a),
\eeq
with $u_a = \sqrt{\frac{J_\lambda}{2}}(a-1)$, eq.\eqref{eq:k0_eq1} becomes
\beq \label{eq:k0_final}
\lambda_d n^+ = \sqrt{2J_\lambda} \tan\left( \frac{\sqrt{2J_\lambda}}{2}(1-a) \right).
\eeq
We recover the asymptotic result \cite{cartailler2017analysis} for positive ions in a ball. The solutions for $n^+ \geq 0$ leads to $0 \leq J_\lambda \leq \frac{\pi^2}{2(1-a)^2}$. The potential difference  is
\beq
u(1) - u(a) = -2 \ln\left(2 \cos\left( \sqrt{\frac{J_\lambda}{2}} (1-a) \right)\right),
\eeq
where $J_\lambda$ is the solution of eq.\eqref{eq:k0_final} for a given $n^+$.
%%%%%%%%%%%%%%%%%%%%%%%%%%%%%%%%%%%%%%%%%%%%%%%%%%%%%%
\subsubsection{Case $N = 0$ ($n^+ = n^-$)}
\begin{figure}[http!] \centering
\includegraphics[scale=0.3]{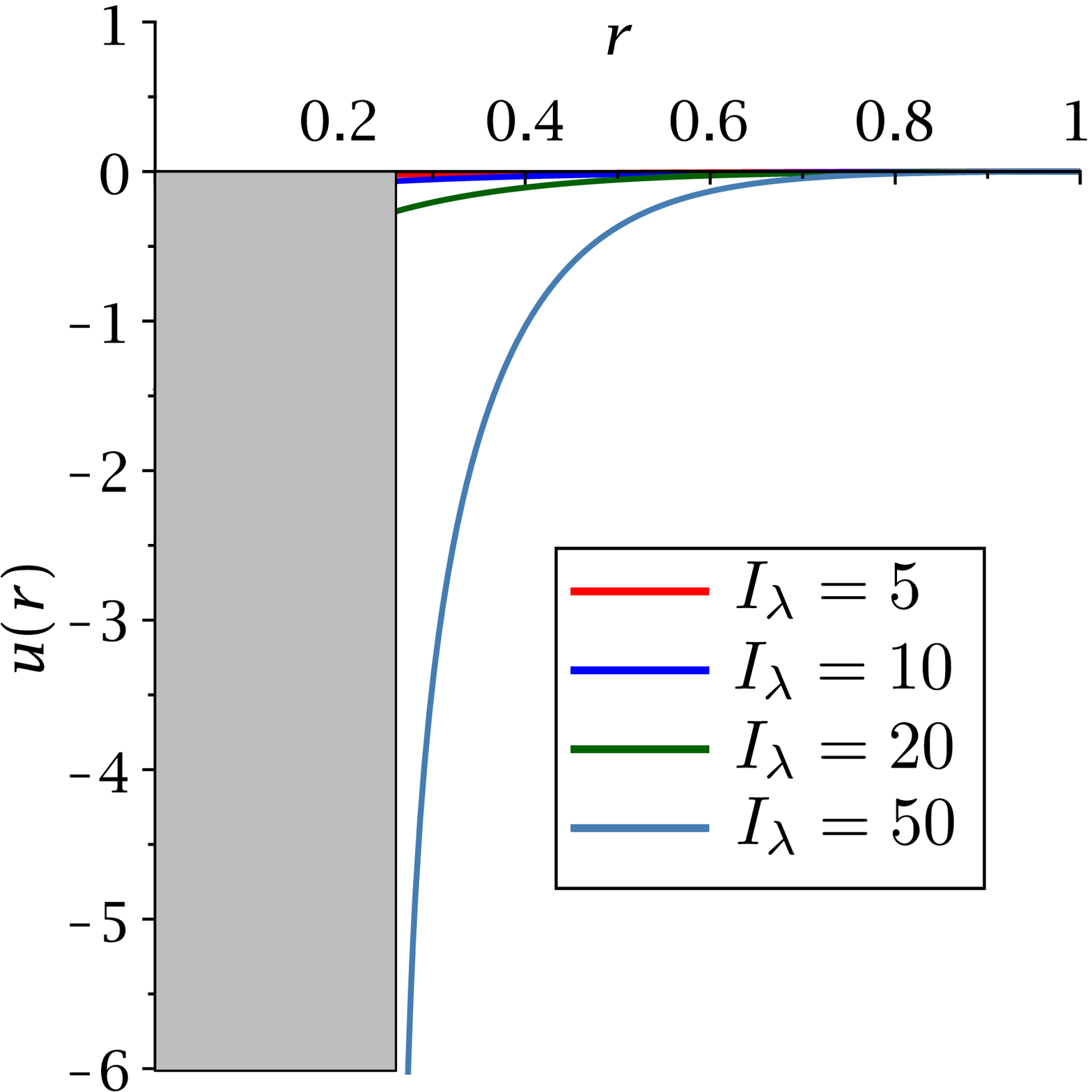}
\includegraphics[scale=0.3]{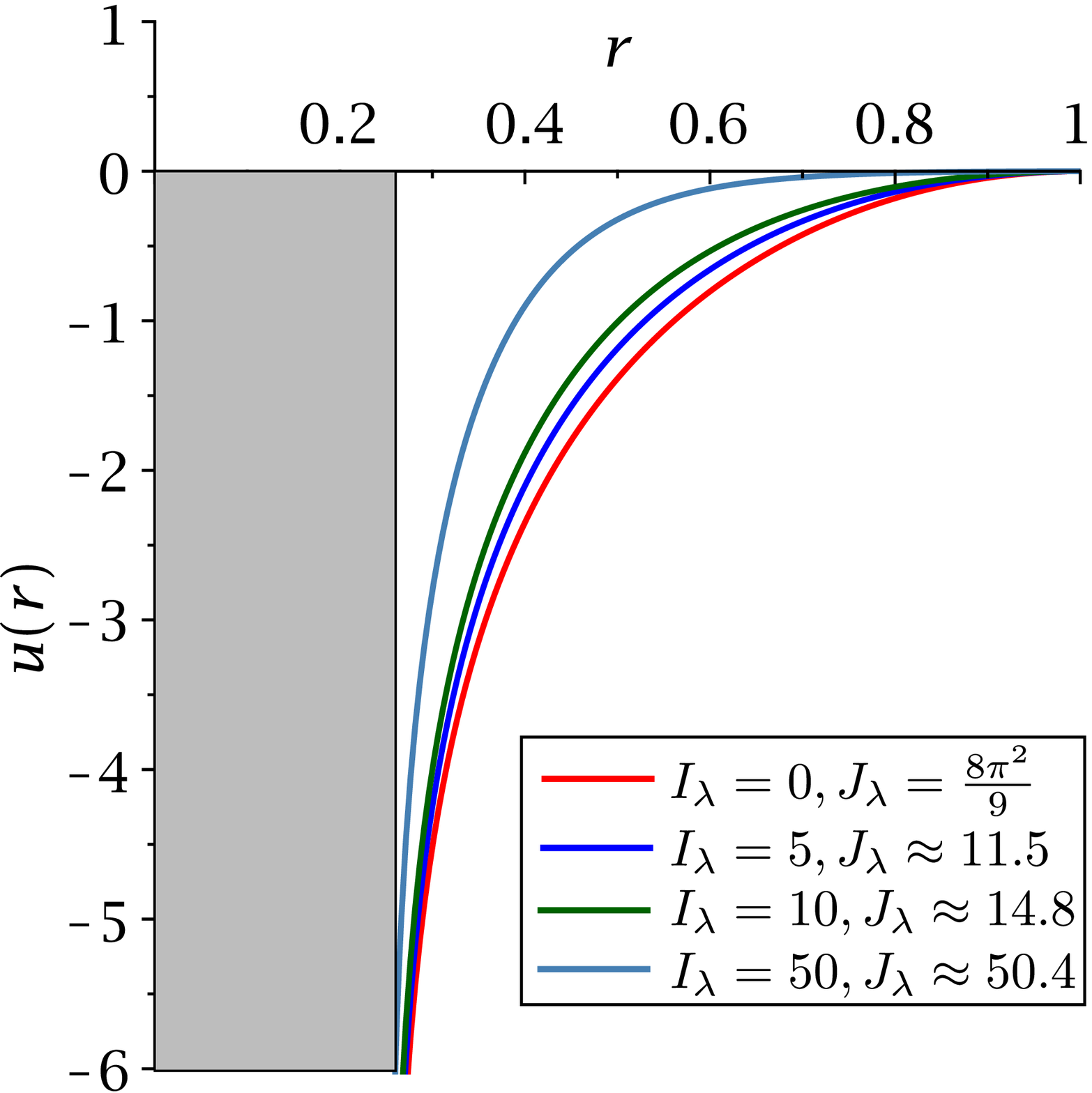}
  \caption{{\bf Left:} Graph of $u$ for $I_\lambda \approx J_\lambda$ and different values of $I_\lambda$.{\bf Right:} Graph of $u$ near the singularity for different values of $I_\lambda$ and $J_\lambda$. $I_\lambda$ and $J_\lambda$ are such that $u_a = -K(k)$.} \label{fig:N0}
%%%%%%%%%%%%%%%%%%%%%%%%%%%%%%%%%%%%%%%%%%%%%%%%%%%%%%%%%%%%%%%%%%%%%%
\end{figure}

%%%%%%%%%%%%%%%%%%%%%%%%%%%%%%%%%%%%%%%%%%%%%%%%%%%%%%
In the case $N = 0$, eq.\eqref{eq:SN} implies $I_\lambda = J_\lambda$ and thus $k_\lambda  = 1$. Eq.\eqref{eq:SJ_appendix1} becomes
\beq \label{eq:k1_eq1}
\lambda_d n^+ = \sqrt{\frac{J_\lambda}{2}} f(u_a).
\eeq
When  $k_\lambda=1$, Jacobian elliptic functions simplify to hyperbolic functions, which gives
$ E(u) = \tanh(u) \ , \quad \operatorname{sn}(u) = \tanh(u) \ , \quad \operatorname{cn}(u) = \operatorname{dn}(u) = \frac{1}{\cosh(u)},$ and $ f(u_a) = 2\tanh(u_a) - u_a -2\tanh(u_a) = -u_a$. Eq.\eqref{eq:k1_eq1} becomes
\beq\label{eq:k1_final}
J_\lambda = \frac{\lambda_d n^+}{1-a},
\eeq
the Jacobian function $\operatorname{dc}=1$, and thus $u(r) = 0$ for $r \in [a,1]$. The behavior of $u$ for small $N$ is shown in fig. \ref{fig:N0} with $J_\lambda = 1.01 I_\lambda$.
%%%%%%%%%%%%%%%%%%%%%%%%%%%%%%%%%%%%%%%%%
%%%%%%%%%%%%%%%%%%%%%%%%%%%%%%%%%%%%%%%%%%
Expanding the Jacobian elliptic functions for $k$ near $1$, we obtain
\beq
 \operatorname{dc}(u) &= 1 + \frac12 (1-k^2) \sinh^2(u) + \circ(1-k^2),\\
 \operatorname{nc}(u) &= \cosh(u) + \circ(1-k^2),
\eeq
and thus
\beqq
u(1) - u(a) &= 2 \ln\left( \frac12 \frac{\sqrt{I_\lambda} + \sqrt{J_\lambda}}{\sqrt{J_\lambda}} \left( 1 + \frac12 (1-k^2)\sinh^2(u_a) + \sqrt{1-k^2} \cosh(u_a)  \right) \right)
%&= 2 \ln\left( \frac12 \frac{\sqrt{I_\lambda} + \sqrt{J_\lambda}}{\sqrt{J_\lambda}} \left( 1 + \sqrt{1-k^2} + \frac12 (1-k^2)\sinh^2(u_a) + \sqrt{1-k^2} \left( \cosh(u_a) -1 \right) \right) \right).
\eeqq
%We have $\sqrt{1-k^2} = \frac{\sqrt{J_\lambda}- \sqrt{I_\lambda}}{\sqrt{J_\lambda} + \sqrt{I_\lambda}}$ and $1+ \sqrt{1-k^2} = \frac{2 \sqrt{J_\lambda}}{\sqrt{I_\lambda} + \sqrt{J_\lambda}}$, so
%\beq
%  u(1) - u(a) = 2 \ln\left( 1 + \frac12 \frac{\sqrt{J_\lambda} - \sqrt{I_\lambda}}{\sqrt{J_\lambda}} \left( \frac12 \sqrt{1-k^2} \sinh^2(u_a) + \cosh(u_a) - 1  \right) \right)
%\eeq
Since $k \longrightarrow 1$, we finally obtain
\beq \label{eq:potential_N0}
u(1) - u(a) \sim \frac{\sqrt{J_\lambda} - \sqrt{I_\lambda}}{\sqrt{J_\lambda}} \left( \frac12 \frac{\sqrt{J_\lambda}- \sqrt{I_\lambda}}{\sqrt{J_\lambda} + \sqrt{I_\lambda}} \sinh^2(u_a) + \cosh(u_a) - 1 \right).
\eeq
%%%%%%%%%%%%%%%%%%%%%%%%%%%%%%%%%%%
\subsubsection{Case $u_a \ll 1$}
%%%%%%%%%%%%%%%%%%%%%%%%%%%%%%%%%%%
For $u_a \ll 1$ we can obtain from Appendix \ref{appendix2} (expression of $f$ and $g$ in eq.\eqref{eq:fg})
\beq
&f(u_a) \sim (k^2 - 2) u_a = \frac{2}{\sqrt{2}} \frac{I_\lambda + J_\lambda}{\sqrt{I_\lambda} + \sqrt{J_\lambda}} (1-a),\\
    &g(u_a) \sim 2 u_a = - \frac{2}{\sqrt{2}} \left( \sqrt{I_\lambda} + \sqrt{J_\lambda} \right) (1-a),
\eeq
so eq.\eqref{eq:SJ_appendix1} gives
\beq
\frac{\lambda_d n^-}{1-a} &= \frac12 \left( I_\lambda + J_\lambda -  \left( J_\lambda - I_\lambda \right)  \right),
\hbox{ and } \frac{\lambda_d n^+}{1-a} &= \frac12 \left( I_\lambda + J_\lambda +  J_\lambda - I_\lambda  \right).
\eeq
For $u_a \ll 1$,
\beq \label{eq:result_small_ua}
\frac{\lambda_d n^-}{1-a} &= I_\lambda \hbox{ and } \frac{\lambda_d n^+}{1-a} &= J_\lambda.
\eeq
Since $u_a = \frac{\sqrt{I_\lambda} + \sqrt{J_\lambda}}{\sqrt{2}} (a-1)$, equations eq.\eqref{eq:result_small_ua} corresponds to few ions. To compute potential difference , we expand the Jacobian elliptic functions $\operatorname{dc}$ and $\operatorname{nc}$ for $u \ll 1$:
\beq
\operatorname{dc}(u) &= 1 + (1-k^2) \frac{u^2}{2} + \circ(u^2), \,\operatorname{nc}(u) &= 1 + \frac{u^2}{2} + \circ(u^2),
\eeq
which gives
\beq
\operatorname{dc}(u_a) + \sqrt{1-k^2} \operatorname{nc}(u_a) &= 1 + \sqrt{1-k^2} + \frac12 \left( 1-k^2 + \sqrt{1-k^2} \right)u^2 + \circ(u^2) \nonumber\\
%&\sim \frac{2\sqrt{J_\lambda}}{\sqrt{I_\lambda} + \sqrt{J_\lambda}}+ \frac12 \left( \frac{\left( \sqrt{J_\lambda} - \sqrt{I_\lambda} \right)^2}{\left( \sqrt{J_\lambda} + \sqrt{I_\lambda} \right)^2} + \frac{J_\lambda - I_\lambda}{\left( \sqrt{J_\lambda} + \sqrt{I_\lambda} \right)^2}\right)u_a^2\\
&\sim \frac{2\sqrt{J_\lambda}}{\sqrt{I_\lambda} + \sqrt{J_\lambda}} + \frac12 \sqrt{J_\lambda}\left(\sqrt{J_\lambda} - \sqrt{I_\lambda}\right) (1-a)^2
\eeq
and thus
\beq
u(1) - u(a) \sim 2 \ln\left(1 + \frac14 \left( J_\lambda - I_\lambda \right) (1-a)^2 \right) \sim \frac12 \lambda_d N (1-a).
\eeq
%When $N \longrightarrow 0$ (eq.\eqref{eq:potential_N0}) and  $u_a \ll 1$,
%this results in the same expression as in eq.\ref{eq:potential_ua0}.
%%%%%%%%%%%%%%%%%%%%%%%%%%%%%%%%
\subsubsection{Case $u_a \longrightarrow -K(k)$}
%%%%%%%%%%%%%%%%%%%%%%%%%%%%%%%%%%%%%%%%%
When $u_a \longrightarrow -K(k)$, we expand with respect to $u_a + K(k)$  the functions $f$ and $g$ using relation eq.\eqref{eq:fg}:
\beq
f(u_a) &= -2E(K(k)) + \left( 2-k^2 \right)K(k) + \frac{2}{u_a + K(k)},\\
g(u_a) &= -\frac{2}{\sqrt{1-k^2}(u_a+K(k))}.
\eeq
From eq.\eqref{eq:SJ_appendix1}, we get
\beq  \label{eq:SIJ_limit}
\lambda_d n^- &=& \frac{\sqrt{I_\lambda} + \sqrt{J_\lambda}}{2\sqrt{2}} \left(-2E(K(k)) + \left( 2-k^2 \right)K(k) \right),\\
\lambda_d n^+ &=& \frac{\sqrt{I_\lambda} + \sqrt{J_\lambda}}{2\sqrt{2}} \left(-2E(K(k)) + \left( 2-k^2 \right)K(k) + \frac{4}{u_a+K(k)}\right),
\eeq
thus $n^+ \longrightarrow \infty$. Using
\beq\label{eq:SN_limit}
\lambda_d (n^+ - n^-) = \lambda_d N = \frac{\sqrt{I_\lambda} + \sqrt{J_\lambda}}{2\sqrt{2}} \frac{4}{u_a+K(k)},
\eeq
we obtain that $N \longrightarrow \infty$. Note that for $k=0$, $E(K(0)) = K(0) = \frac{\pi}{2}$ so $n^- = 0$. However, when $k \longrightarrow 1$, $K(k) \longrightarrow \infty$, then $n^- \longrightarrow \infty$. The singularity is located at $r = a - \varepsilon$ with $\varepsilon \ll 1$. Since $ u_a = \frac{\sqrt{I_\lambda} + \sqrt{J_\lambda}}{\sqrt{2}}(a-1)$, we have
\beq
u_{a-\varepsilon} = \frac{\sqrt{I_\lambda} + \sqrt{J_\lambda}}{\sqrt{2}}(a-\varepsilon-1) = - K(k)
\eeq
and
\beq \label{eq:ua_epsilon}
u_a = -K(k) + \frac{\sqrt{I_\lambda} + \sqrt{J_\lambda}}{\sqrt{2}} \varepsilon.
\eeq
Expanding the Jacobian elliptic functions $\operatorname{nc}$ and $\operatorname{dc}$ near $-K(k)$ :
\beq
\operatorname{dc}(u) \sim \frac{1}{u + K(k)} , \qquad \operatorname{nc}(u) \sim \frac{1}{\sqrt{1 - k^2}(u + K(k))},
\eeq
 using eq.\eqref{eq:potential_dim1} and eq.\eqref{eq:ua_epsilon},we obtain
\beq
u(1) - u(a) = 2 \ln\left( \sqrt{\frac{2}{J_\lambda}} \frac{1}{\varepsilon} \right).
\eeq
From the expression of $u_a + K(k)$ in eq.\eqref{eq:ua_epsilon} and the formula for $\lambda_d N$ in eq.\eqref{eq:SN_limit}, we get $\lambda_d N = \frac{2}{\varepsilon}$ and finally
\beq\label{eq:potential_dim1_limit}
u(1) - u(a) = 2 \ln\left( \frac{\lambda_d N}{\sqrt{2 J_\lambda}} \right).
\eeq
When $n^- = 0$, similar to section \ref{n-0}, we get
\beq
\lambda_d n^+ = \sqrt{2J_\lambda} \tan\left( \frac{\sqrt{2J_\lambda}}{2}(1-a) \right)
\eeq
and since $n^+ \longrightarrow \infty$, $\sqrt{2 J_\lambda} \longrightarrow \frac{\pi}{1-a}$, we finally get
\beq
u(1) - u(a) \sim 2\ln\left( \frac{(1-a)\lambda_d N}{\pi} \right).
\eeq
If $n^- \longrightarrow \infty$, from eq.\eqref{eq:SIJ_limit} $K(k) \longrightarrow \infty$,
which means that $k \longrightarrow 1$ and thus $J_\lambda - I_\lambda \longrightarrow 0$.
We can make the approximation
\beq
  \sqrt{I_\lambda} + \sqrt{J_\lambda} \approx 2 \sqrt{J_\lambda}
\eeq
and write
\beqq
\lambda_d n^- &= \frac{\sqrt{I_\lambda} + \sqrt{J_\lambda}}{2\sqrt{2}} \left(-2E(K(k)) + \left( 2-k^2 \right)K(k) \right) &\sim \frac{\sqrt{I_\lambda} + \sqrt{J_\lambda}}{2\sqrt{2}} K(k)
  %  \\
%    &\sim \frac{\sqrt{I_\lambda} + \sqrt{J_\lambda}}{2\sqrt{2}} (-u_a)
%    \\
%    &\sim \frac{\left(\sqrt{J_\lambda} + \sqrt{I_\lambda}\right)^2}{4} (1-a).
%    \\
\sim J_\lambda (1-a)
\eeqq
Then using the expression of the potential difference  in eq.\eqref{eq:potential_dim1_limit}, we obtain
\beq \label{ddpfinal}
u(1) - u(a) \sim 2\ln\left(\sqrt{\lambda_d (1-a)} \frac{N}{\sqrt{n^-}} \right).
\eeq
In particular, when $n^- = N$, we get
\beq
u(1) - u(a) \sim \ln\left(\lambda_d N (1-a) \right).
\eeq
We have also plotted the function normalized potential $u(r)$ in Fig. \ref{fig:N0}-Right.
%%%%%%%%%%%%%%%%%%%%%%%%%%%%%%%%%%%%%%%%%%%%%%%%%%
\subsection{Summary potential difference  $u(1)-u(a)$}
\begin{figure}[http!]
  \centering
  \includegraphics[scale=0.35]{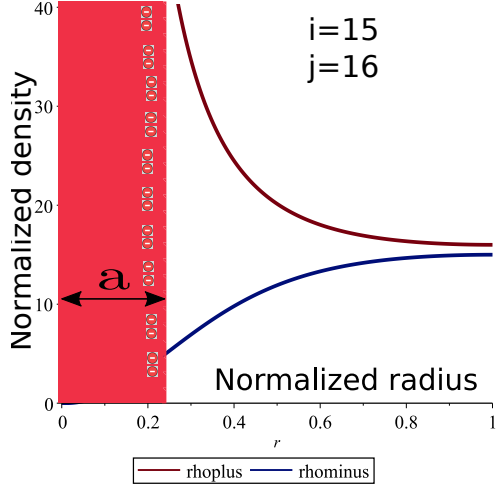}
  \caption{Distribution of positive and negative charge for $I_{\lambda}=i=15$ and $J_{\lambda}=j=16$ and $a=0.25$ associated to $n^+ \lambda_d= 9.308$ and $n^- \lambda_d= 15.58$.}
  \label{fig:distributioncharge}
\end{figure}

%%%%%%%%%%%%%%%%%%%%%%%%%%%%%%%%%%%%%%%%%%%%%%%%%%
We summarize in the table \ref{table:sol} below the differences of potential $u(1)-u(a)$ for the explicit solution in dimension 1, depending on the different condition on the mobile positive $n^+$ and negative $n^-$ charges satisfying the global electro-neutrality conditions $N+n^-=n^+$.
%%%%%%%%%%%%%%%%%%%%%%%%%%%%%%%%%%%%%%%%%%%%%%%%%%
\begin{table}[H]
\begin{center}
\begin{tabular}{|c|c|  }
\hline Conditions & $  u(1)-u(a)$   \\ \hline
& \\
$\begin{matrix}
n^- \longrightarrow 0(n^+ = N) \\
\lambda_d n^+ = \sqrt{2J_\lambda} \tan\left( \frac{\sqrt{2J_\lambda}}{2}(1-a) \right)\\
\end{matrix}$&
$\ds-2 \ln\left(2 \cos\left( \sqrt{\frac{J_\lambda}{2}} (1-a) \right)\right)$
     \\
	& \\
$\begin{matrix}
N = 0 (n^+ \sim n^-)\\
I_\lambda = J_\lambda\\
J_\lambda = \frac{\lambda_d n^+}{1-a}\\
u_a=\sqrt{2I_\lambda} (a-1)\\
\end{matrix}$&
$\ds  \frac{\sqrt{J_\lambda} - \sqrt{I_\lambda}}{\sqrt{J_\lambda}} \left( \frac12 \frac{\sqrt{J_\lambda}- \sqrt{I_\lambda}}{\sqrt{J_\lambda} + \sqrt{I_\lambda}} \sinh^2(u_a) + \cosh(u_a) - 1 \right).$ \\
	& \\
	$\begin{matrix}
	\frac{\sqrt{I_\lambda} + \sqrt{J_\lambda}}{\sqrt{2}} (a-1) \ll 1\\
	\frac{\lambda_d n^-}{1-a} = I_\lambda  \\
	\frac{\lambda_d n^+}{1-a} = J_\lambda
	\end{matrix}$ & $\ds 2 \ln\left(1 + \frac14 \left( J_\lambda - I_\lambda \right) (1-a)^2 \right) \sim \frac12 \lambda_d N (1-a)$   \\
	 &  \\
$\begin{matrix}
N,n^- ,n^+ \gg1\\
\end{matrix}$&
$\ds  2\ln\left(\sqrt{\lambda_d (1-a)} \frac{N}{\sqrt{n^-}} \right).$ \\
	& \\
$\begin{matrix}
N, n^+ \gg1\\
n^-=0\\
\end{matrix}$&
$\ds  2\ln\left( \frac{(1-a)\lambda_d N}{\pi} \right).$ \\
	& \\
%	$\begin{matrix}
%    \tilde\sigma_{\eps}<0\,,\,\tilde\sigma_{cusp}\sqrt{\eps}\gg1\\
%    \tilde\sigma_{bulk}>0\\
%    (\mbox{2D})
%	\end{matrix}$  & {\small $\ds\frac{k T}{e}\left (-2\ln\frac{\sqrt{2}\,e\pi
%			R_c\tilde{\sigma}_{cusp} }{   4kT  +  e\tilde \eps\tilde {\sigma}_{cusp}   } +
%	\frac{ e\tilde \eps\tilde{\sigma}_{\eps}}{kT}  \left(\frac{2\tilde{\sigma}_{cusp}-\tilde{\sigma}_{\eps}}{\tilde{\sigma}_{cusp}}  \right) \right)$}   \\
%	&  \\
	\hline
\end{tabular}
\caption{Electrodiffusion relations for the potential difference. \label{table:sol}}
\end{center}
\end{table}
%%%%%%%%%%%%%%%%%%%%%%%%%%%%%%%%%%%%%%%%%%%%%%%%%%
Finally, in fig. \ref{fig:distributioncharge} we show the distribution of positive (red) and negative (blue) charge density computed in dimension 1 inside $[a,1]$, associated to  $u(a)-u(1)=7.207$ and $\lambda_d N=0.0887$. Note that the difference of charge persists deep inside the domain.
%%%%%%%%%%%%%%%%%%%%%%%%%%%%%%%%%%%%%%%5
%%%%%%%%%%%%%%%%%%%%%%%%%%%%%%%%%%%%%%%%5
\newpage
%%%%%%%%%%%%%%%%%%%%%%%%%%%%%%%%%%%%%%%%%%%%%%%%%%%%%%
\section{Steady-solution in two dimensions}\label{ss:SSS3}
%%%%%%%%%%%%%%%%%%%%%%%%%%%%%%%%%%
In this section, we resolve the PNP equation \ref{eq:pnpgeneral} in two dimensions (Fig. \ref{fig:pnpdim2_bloc}A), which reduces to
\beq \label{eq:pnpdim2}
u''(r) + \frac1r u'(r) = I_\lambda e^{u(r)} - J_\lambda e^{-u(r)},
\eeq
with the boundary conditions
\beq \label{eq:pnpdim2_boundary}
  u(1) = u'(1) = 0.
\eeq
We first solve this equation when there are no moving negative ions (Fig. \ref{fig:pnpdim2_bloc}B-C) and then use a regular perturbation to find the general solution.
%%%%%%%%%%%%%%%%%%%%%%%%%%%%%%%%%%%%%%%%%%
\subsection{No negative ions : $I_\lambda = 0$}
%%%%%%%%%%%%%%%%%%%%%%%%%%%%%%%%%%%%%%%%%%
In the new variables
\beq
r &= e^{-t}\, \tilde{u}(t) &= u(r) + 2t.
\eeq
eq. \eqref{eq:pnpdim2} is transformed into
\beq
\tilde{u}''(t) = - J_\lambda e^{-\tilde{u}(t)},
\eeq
with boundary conditions $ \tilde{u}(0) = 0, \quad \tilde{u}'(0) = 2.$ A first integration gives
\beq  \label{eq:dim2_int}
\frac12 \tilde{u}'^2 = J_\lambda e^{-\tilde{u}(t)} + 2 - J_\lambda.
\eeq
There are three cases: $J_\lambda < 2$, $J_\lambda = 2$ and $J_\lambda > 2$ we show in appendix \ref{appendix4} the following explicit solutions
\beq \label{summarydim2}
J_\lambda < 2 \quad : \quad u_0(r) &=& 2 \ln\left( \frac12 \left( 1 + \frac1p \right) r^{1-p} -\frac12 \left( \frac1p - 1 \right)r^{1+p} \right), \quad p = \sqrt{1-\frac{J_\lambda}{2}} \nn\\
J_\lambda = 2 \quad : \quad u_0(r) &=& 2 \ln(r(1-\ln(r))) \\
J_\lambda > 2 \quad : \quad u_0(r) &=& 2 \ln\left( r \left( \frac1p \sin(-p \ln(r)) + \cos(-p \ln(r)) \right) \right), \quad p = \sqrt{\frac{J_\lambda}{2}-1}. \nn
\eeq
%%%%%%%%%%%%%%%%%%%%%%%%%%%%%%%%%%%%%%%%%%%%%%%%%%
\subsubsection{Regular perturbation solution for $I_\lambda=\varepsilon \ll1$}
\begin{figure}[http!]
\centering
\includegraphics[width=\textwidth]{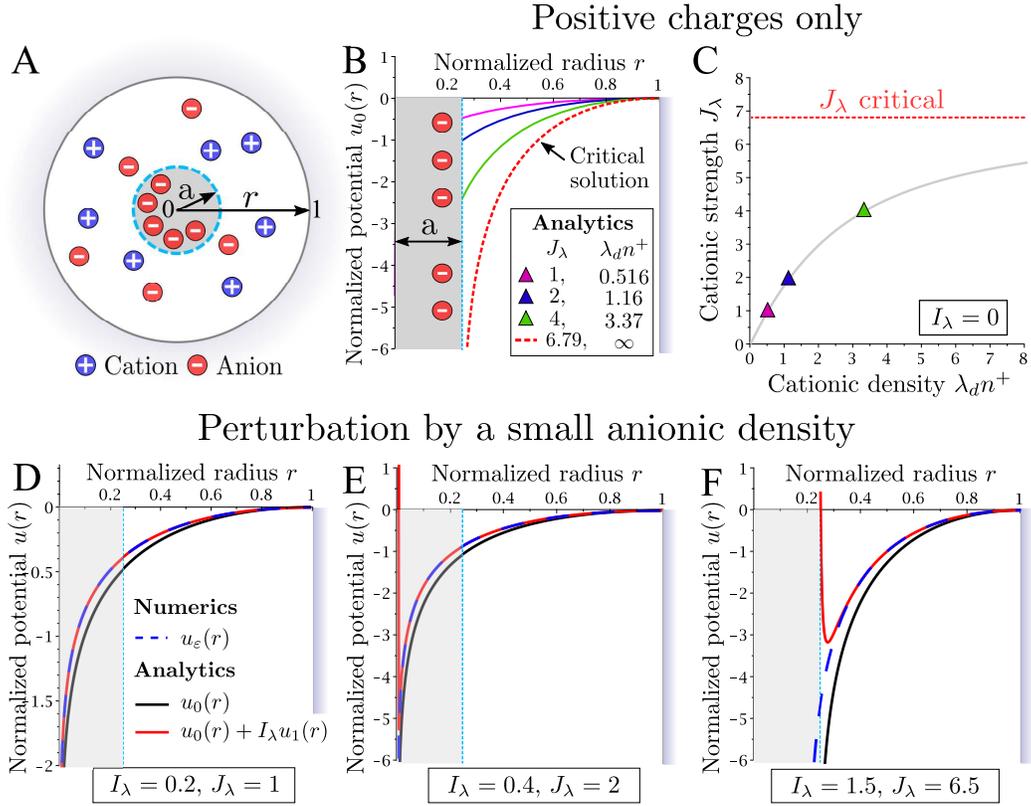}
\caption{Normalized voltage distribution in two dimensions. \textbf{A.} Scheme of domain and charge distribution in the annulus.
\textbf{B,C.} No negative ions are present in the region $[R_0=0.25,R=1]$, \textbf{B.} Solution for different values of $I_\lambda$ and \textbf{C.} $J_\lambda$ vs the cationic density $\lambda_d n^+$. \textbf{D-F.} Approximated solution computed from the regular expansion (eq. \ref{eq:dim2_uepsilon}) compared to the exact solution computed numerically. }
\label{fig:pnpdim2_bloc}
\end{figure}

\begin{figure}[http!]
  \centering
  \includegraphics[scale=0.35]{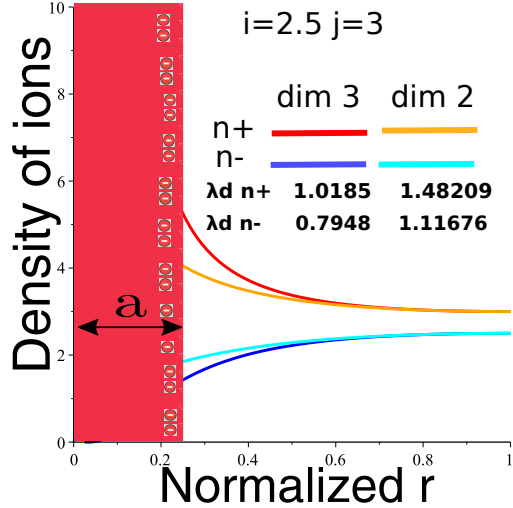}
  \caption{Distribution of positive and negative charges for $i=2.5$ and $j=3$ and $a=0.25$ in dimension 2 and 3. Positive (resp. negative) charges in dimension 3 (red, resp. blue) and dimension 2 (orange resp. cyan).}
  \label{fig:distributioncharge2d3d}
\end{figure}

%%%%%%%%%%%%%%%%%%%%%%%%%%%%%%%%%%%%%%%%%%%%%%%%%%%%%%%%%%%%%
We expand the solution $u_\varepsilon = u_0 + \varepsilon u_1 + \circ(\varepsilon)$, where $u_\varepsilon$ is the solution of
\beq  \label{eq:dim2_uepsilon}
u_\varepsilon''(r) + \frac1r u_\varepsilon'(r) = \varepsilon e^{u_\varepsilon(r)} - J_\lambda e^{-u_\varepsilon(r)} \\
u_\varepsilon(1) = u_\varepsilon'(1) = 0,
\eeq
where $u_0$ is given in \ref{summarydim2} and $u_1$ satisfies:
\beq \label{eq:dim2_u1}
u_1''(r) + \frac1r u_1'(r) = e^{u_0(r)} + J_\lambda e^{-u_0(r)} u_1(r),
 \eeq
with the initial conditions
\beq
  u_1(1) = u_1'(1) = 0.
\eeq
We now discuss the solution in the three cases $J_\lambda < 2$, $J_\lambda = 2$ and $J_\lambda > 2$.
For $J_\lambda = 2$, the solution of eq.\eqref{eq:dim2_u1} is
\beq
u_1(r) = \left( A + \lambda(r) \right) \left( 1 - \ln(r) \right)^2 + \frac{B + \mu(r)}{1-\ln(r)},
\eeq
where
\beq
A=\frac{5}{48}, \, B= -\frac{103}{384} \, \lambda(r) = \frac{r^4}{12} \left( -\frac{5}{4} + \ln(r) \right)\\
\mu(r) = \frac{r^4}{384} \left( 32 \ln(r)^4 - 160 \ln(r)^3 + 312 \ln(r)^2 - 284 \ln(r) + 103 \right).
\eeq
In the cases $J_\lambda < 2$ and $J_\lambda > 2$, we use numerical simulations to estimate the perturbation $u_1$ and plotted in Fig. \ref{fig:pnpdim2_bloc} the normalized voltage obtained numerically and using expansion \ref{eq:dim2_uepsilon}. We found a very good agreement between the numerical and the approximation solutions for  $J_\lambda \leq 2$ in the entire domain (Fig. \ref{fig:pnpdim2_bloc}D-E). However, for $J_\lambda \leq 2$, the approximation diverged from the numerical solution near the boundary of the inner domain ($r=0.25$), Fig. \ref{fig:pnpdim2_bloc}F).
%%%%%%%%%%%%%%%%%%%%%%%%%%%%%%%%%%%%%%%%%
%%%%%%%%%%%%%%%%%%%%%%%%%%%%%%%%%%%%%%%%%
Finally, we show in Fig. \ref{fig:distributioncharge2d3d} the distribution of positive and negative charges.
%%%%%%%%%%%%%%%%%%%%%%%%%%%%%%%%%%%%%%%5
%%%%%%%%%%%%%%%%%%%%%%%%%%%%%%%%%%%
%%%%%%%%%%%%%%%%%%%%%%%%%%%%%%%%%%
\section{Numerical evaluation  of the voltage distribution in three dimensions} \label{sec:ap6}
\begin{figure}[http!]
  \centering
  \includegraphics[scale=0.35]{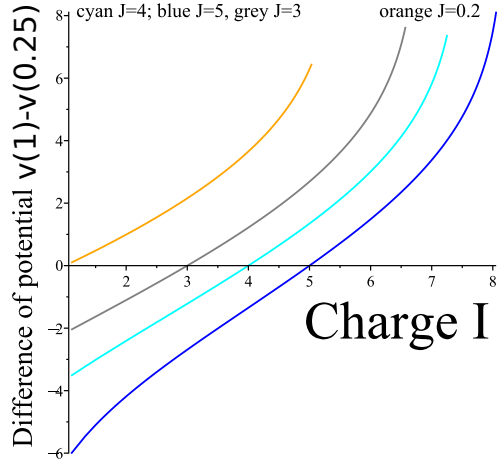}
  \caption{Normalized Potential difference  vs the charge $I_\lambda$ for various negative charge $J_\lambda=0.2$ orange; 4 cyan; 3 grey, 5 cyan) .}
  \label{fig:diffpotntial}
\end{figure}

%%%%%%%%%%%%%%%%%%%%%%%%%%%%%%%%%%
In three dimensions,  eq.\eqref{eq:pnpgeneral} becomes
\beq \label{eq:pnpgeneral3}
 u''(x) + \frac{2}{x} u'(x) = I_\lambda e^{u(x)} - J_\lambda e^{-u(x)},
\eeq
which does not have a direct solution. We solved numerically eq. \ref{eq:pnpgeneral3} with boundary conditions \ref{eq:bc} (Fig. \ref{fig:diffpotntial}).
%%%%%%%%%%%%%%%%%%%%%%%%%%%%%%%%%%%%%%%5
%%%%%%%%%%%%%%%%%%%%%%%%%%%%%%%%%%%
%Finally, we show some potential difference
%%%%%%%%%%%%%%%%%%%%%%%%%%%%%%%%%
\section{Discussion and concluding remarks}
%%%%%%%%%%%%%%%%%%%%%%%%%%%%%%%
We have studied in this article the distribution of the voltage field in a global but non-local electroneutral electrolyte. We found that the voltage does not decay quickly, but quite slowly inside the bulk region due to the local charge imbalance. We could completely resolve the electrodiffusion equations in dimension one (flat geometry) and partially in dimension 2 (cylindrical) using a regular perturbation around the solution with positive ions only and a negative charge in a disk. The solution in dimension three could only be estimated numerically. In all three dimensions, the potential difference between the inner and outer surfaces of the electrolyte should depend on the log of the charges, as we have shown in dimensions 1 and 2 (see table \ref{table:sol}). It will be interesting to extend the present analysis to the case of non-concentric disk and in particular to examine the situation where the inner and outer boundaries could be very close. We also expect that curved membrane will create voltage drops, as shown in \cite{cartailler2017electrostatics} the case of global non-electroneutrality. \\
In many biological nanodomains, such as inside dendritic spines, the concentration of mobile chloride ions is not counterbalanced by the mobile positive ions (potassium, sodium and free calcium ions essentially). For a total of 150 $mM$ positive, the mobile ions are divided into $~18$ mM $Na+$, $~135 mM K+$ and $~0.0001 mM Ca2+$ and $~7 mM Cl-$ ions and it is expected that most negative charges are located in membranes and consist of almost immobile macromolecules. These differences in ion mobility might result in important junction potentials (that is, local depletions in specific ion species), especially during transient synaptic activation, following an important influx of positive charges through AMPA-type glutamate receptors. In the present model, if we consider $n_+=150 mM$  $n_-=7 mM$  $N=143$ in a ball of $1 \mu m$ and an inner domain of $100nm$, then using the dimension 1 approximation for $N\approx 10^8, n+\approx 9.10^7 \gg1$ and $n-\approx 44.10^5 $, we have from eq. \ref{ddpfinal} that
\beq \label{ddpfinal2}
u(1) - u(a) \sim 2\ln\left(\sqrt{\lambda_d (1-a)} \frac{N}{\sqrt{n^-}} \right)=0.167
\eeq
where $\lambda_d=6.97*10^{-10}$, $a=0.25$ and the parameters are given in table \ref{table2}. Thus the voltage difference is $\Delta V=4.33 mV$ in a region of length 750nm.\\
%%%%%%%%%%%%%%%%%%%%%%%%%%%%%%%%%%%%%%%%%%%%%%%%%%%%%%
\begin{table}[h!]
\caption{Parameters.}
\begin{center}
%\scalebox{1.0}
\begin{tabular}{l l l l} %\hline\\
\hline
 \hline  Parameter & \, \, \, \, Description & Value &\\
 %\hline\\
$z$& Valence of ion & z=1 (for sodium)\\
$\Omega$ &  Spine head  & $\Omega$ (volume $|\Omega|=1\mu m^3$) \\
$a$&  size of the negative charge region& (typical) $a=0.25\mu m$\\
$R$&  radius of spine  head & (typical) $L=1\mu m$\\
$T$&  Temperature  &   $T=300K$\\
$E$&  Energy  &   $ kT=2.58\times10^{-2}eV$\\
$e$&  Electron charge  &   $e=1.6\times10^{-19}C$\\
$\eps$&  Dielectric constant  &   $\eps=80$\\
$\eps_0$&  Dielectric constant  &   $\eps=8.85*10^{-12}F/m$\\
%$\Lambda_H$ &  Dimensionless spine head constant  & $\Lambda_H=\ds\frac{|\Omega|zeI}{16kTDa^2\eps\eps_0}\approx17846$.\\
%$\lambda_H$ &  Inverse Spine head impedance &
%$\lambda_H=\ds\frac{|\Omega|ze}{16kTDa^2\eps\eps_0}\approx 7138.4
%*10^{12}$.\\
\hline\\
\end{tabular}
%}
\end{center}\label{table2}
\end{table}
%%%%%%%%%%%%%%%%%%%%%%%%%%%%%%%%%%%%%%%%%%%%%%%%%%%%%%
Finally, this study pushes to test the spatial limit of the electro-neutrality hypothesis in neuronal cell. When a large amount of negatively charged proteins are distributed in a confined microdomain, it would be interesting to investigate the consequences on the regulation of positive ionic distribution entering through channels. In particular, we expect from the present study that injecting a current in a cell when electroneutrality is not satisfied at a scale of $10$ to $100nm$, will lead to long penetrating voltage drop inside the bulk.\\  After sodium positive ions enter a dendritic spine, other positive potassium ions could be expelled quickly, a process that would not happen if positive and negative charges would enter at the same time. A transient entry of positive ions in a non-electroneutrality medium could thus generate an electric field much further away compared to an electroneutrality medium, possibly responsible for the fast propagation of opening and closing of channels along dendrites and axons, a mechanism that could also challenge the classical Hodgkin-Huxley paradigm.
%\newpage
%%%%%%%%%%%%%%%%%%%%%%%%%%%%%%%%%
\section{Appendices}
%%%%%%%%%%%%%%%%%%%%%%%%%%%%%%%%%
\subsection{Direct integration}\label{appendix1}
%\addcontentsline{toc}{section}{\nameref{sec:ap1}}
We solve eq. \ref{eq:bc_dim1} by a direct integration after multiplying by $u'(x)$
equation
\beq
u''(x) = I_\lambda e^{u(x)} - J_\lambda e^{-u(x)} \qquad \text{for $a < x < 1$},
\eeq
we get
\beq
\frac12 u'^2(x) =  I_\lambda e^{u(x)} + J_\lambda e^{-u(x)},
\eeq
where we used the boundary conditions
\beq \label{eq:bc_appendix}
u(1) = 0, \quad u'(1) = 0.
\eeq
We now set $u(x) = v(x) + D$ with $D = \frac12 \ln \left( \frac{J_\lambda}{I_\lambda} \right) \geq 0$ and get
\beq \label{eq:pnpdim1}
v'^2(x) = A (\cosh(v(x)) + c),
\eeq
where
\beq\label{eq:ac}
A = 4\sqrt{I_\lambda J_\lambda}\quad , \quad c = -\frac{I_\lambda + J_\lambda}{2\sqrt{I_\lambda J_\lambda}}.
\eeq
In order to integrate eq.\eqref{eq:pnpdim1}, we compute the integral
\beq \label{eq:intcosh}
I(v) = \int^{v} \frac{du}{\sqrt{\cosh(u) + c}} = \int^v \frac{du}{\sqrt{2 \cosh^2(\frac{u}{2}) - 1 + c}}.
\eeq
Changing the variable $x = \frac{1}{\cosh(\frac{v}{2})}$, we transform integral \ref{eq:intcosh} into
\beq
I(v) %&= -2 \bigints^{\ddfrac{1}{\cosh\left(\frac{v}{2}\right)}} \ddfrac{dx}{\sqrt{x^2(1-x^2)\left(\frac{2}{x^2} - 1 + c\right)}} \\&
= -2 \bigints^{\ddfrac{1}{\cosh\left(\frac{v}{2}\right)}} \ddfrac{dx}{\sqrt{(1-x^2)\left(2 - (1 - c)x^2\right)}}.
\eeq
We define
\beq \label{eq:k}
k = \sqrt{\frac{2}{1-c}} = \ddfrac{2\sqrt[4]{I_\lambda J_\lambda}}{\sqrt{I_\lambda} + \sqrt{J_\lambda}}, \qquad 0 < k \leq 1
\eeq
and set $x = kt$ to obtain the incomplete elliptic integral of the first kind of amplitude $\frac{1}{k cosh\left(\frac{u}{2}\right)}$ and modulus $k$ (eq.\eqref{eq:k} leads to $0 < k \leq 1$):
\beq
I(v) = -\sqrt{2} k \bigints^{\ddfrac{1}{k \cosh\left(\frac{v}{2}\right)}} \ddfrac{dt}{\sqrt{(1 - k^2 t^2)(1 - t^2)}}=K\left(\frac{1}{k \cosh\left(\frac{v}{2}\right)},k\right).
\eeq
Thus from eq.\eqref{eq:pnpdim1}, we get
\beq
K\left(\frac{1}{k \cosh\left(\frac{v(x)}{2}\right)},k\right) = \alpha - \sqrt{\frac{A}{2}}\frac{x}{k},
\eeq
where $\alpha$ is a constant. Since $u(1) = 0$, $v(1) = -D$ thus $\cosh\left(\frac{v(1)}{2}\right) = \frac{1}{k}$ and $\alpha = K(1,k) + \sqrt{\frac{A}{2}}\frac{1}{k}$. Using the Jacobian elliptic functions of modulus $k$, we obtain the explicit expression for $v$ with respect to $x$ using the identity $K(.,k) = \operatorname{sn}^{-1}(.)$. Finally,
\beq
\frac{1}{k \cosh\left(\frac{v(x)}{2}\right)} = \operatorname{sn}\left( K(1,k) + \sqrt{\frac{A}{2}}\frac{1-x}{k}\right) = \operatorname{cd} \left( \sqrt{\frac{A}{2}}\frac{x-1}{k} \right),
\eeq
and $v(x) \leq 0$ for $a \leq x \leq 1$,
\beq\label{eq:dim1v}
v(x) = -2 \operatorname{arcosh}\left[ \frac{1}{k} \operatorname{dc}\left( \sqrt{\frac{A}{2}}\frac{x-1}{k} \right) \right].
\eeq
In the following part, we will write $K(k) = K(1,k)$ the complete elliptic integral of the first kind. The normalize potential is
{\small \beq  \label{eq:dim1_u_appendix}
u(x) = -2 \ln\left( \frac12 \frac{\sqrt{I_\lambda} + \sqrt{J_\lambda}}{\sqrt{J_\lambda}} \left( \operatorname{dc}\left(\frac{\sqrt{I_\lambda} + \sqrt{J_\lambda}}{\sqrt{2}} (x-1) \right) + \sqrt{1 - k^2} \operatorname{nc}\left(\frac{\sqrt{I_\lambda} + \sqrt{J_\lambda}}{\sqrt{2}} (x-1) \right) \right) \right).
\eeq
}
%%%%%%%%%%%%%%%%%%%%%%%%%%%%%%%%%%%%%%%%%%%%%%%%%%%%%%%%%%%%%%%%%%%%%%%%%%%%%%
\subsection{Appendix 2: classical relations between elliptic functions}
\label{appendix2}
%\addcontentsline{toc}{section}{\nameref{sec:ap2}}
The incomplete elliptic integral of the first kind of modulus $k$ and argument $x$ is defined by
\beq
K(x,k) = \int_0^\phi \frac{d\theta}{\sqrt{1 - k^2 \sin^2(\theta)}} =\int_0^x \frac{dt}{\sqrt{(1-t^2)(1-k^2 t^2)}},
\eeq
where $x=\sin \phi$. For $x=1$, we obtain the complete elliptic integral of modulus $k$:
\beq
K(k) = \int_0^1 \frac{dt}{(1-t^2)(1-k^2 t^2)}
\eeq
The \textit{elliptic sine $sn$} of modulus $k$ and the \textit{elliptic cosine $cn$} of modulus $k$ are defined by
\beq
sn(K(x,k),k) = \sin \phi = x, \quad cn(K(x,k),k) = \cos \phi.
\eeq
We shall omit the $k$ argument so that $sn(u,k)=sn(u)$. The \textit{delta amplitude} is defined by
\beq
dn(u) = \sqrt{1 - k^2 sn(u)}.
\eeq
The other nine Jacobian elliptic functions are obtained as ratios of the three first ones, following the formula
\beq
pq(u) = \frac{pn(u)}{qn(u)},
\eeq
where $p$ and $q$ are any of the letter $n$,$s$,$c$,$d$, and $nn(u) = 1$.
For example,
\beq
sc(u) = \frac{sn(u)}{cn(u)} \quad \text{and} \quad nc(u) = \frac{1}{cn(u)}.
\eeq
Squares of the functions are obtained from the two relations :
\beq
sn^2(u) + cn^2(u) &= 1,\\(1-k^2) sn^2(u) + cn^2(u) &= dn^2(u).
\eeq
%%%%%%%%%%%%%%%%%%%%%%%%%%%%%%%%%%%%%%%%
%\begin{figure}[http!]
%\centering
% \includegraphics[width=\textwidth]{jacobi.eps}
% \caption{Behaviour of the Jacobian elliptic functions $sn$, $cn$, $dn$ for $k=\frac12$. The x-axis goes from $0$ to $4 K(k)$}
%  \label{fig:jacobi}
%\end{figure}
%%%%%%%%%%%%%%%%%%%%%%%%%%%%%%%%%%%%%%%%
%%%%%%%%%%%%%%%%%%%%%%%%%%%%%%%%%%%%%%%
\subsection{Relations between parameters $I_\lambda$ and $J_\lambda$} \label{appendix3}
%\addcontentsline{toc}{section}{\nameref{sec:ap3}}
We provide here expressions between $I_\lambda$ and $J_\lambda$:  since $\operatorname{arcosh}(x) = \ln\left( x + \sqrt{x^2 - 1} \right)$ for $x \geq 1$, we have
\beq \label{eq:integral1}
e^{-v(x)} = \left( \frac{1}{k} \operatorname{dc}\left( \sqrt{\frac{A}{2}}\frac{x-1}{k} \right) + \sqrt{\frac{1}{k^2} \operatorname{dc}^2 \left( \sqrt{\frac{A}{2}}\frac{x-1}{k} \right) - 1} \right)^2.
\eeq
Using the modulus $k$ of the Jacobian elliptic function $dc$, we have $dc^2(u) - k^2 = (1 - k^2) \operatorname{nc}^2(u)$ and then
\beq
e^{-v(x)} = \left( \frac{1}{k} \operatorname{dc}\left( \sqrt{\frac{A}{2}}\frac{x-1}{k} \right) + \frac{\sqrt{1 - k^2}}{k} \operatorname{nc}\left( \sqrt{\frac{A}{2}}\frac{x-1}{k} \right) \right)^2.
\eeq
We expand this expression and use the following integrals
\beq \label{eq:square_jacobis}
&\int^{u} \operatorname{dc}^2(x) dx = -E(u) + u + \operatorname{sn}(u) \operatorname{dc}(u),\\
&\int^{u} \operatorname{nc}(x) \operatorname{dc}(x) dx = \operatorname{sc}(u),\\
\left( 1 - k^2 \right) &\int^{u} \operatorname{nc}^2(x) dx = -E(u) + \left(1-k^2\right) u + \operatorname{sn}(u) \operatorname{dc}(u),
\eeq
where $E$ is the incomplete elliptic integral of the second kind of modulus $k$,
\beq
E(u) = \int_{0}^{sn(u)} \sqrt{\frac{1 - k^2 x^2}{1 - x^2}} dx.
\eeq
This leads to
\beq \label{eq:int_jlambda}
&\int_{a}^{1} e^{-v(x)} dx = \frac{1}{k}\sqrt{\frac{2}{A}} \left( 2 \operatorname{E}(u_a) - 2 \operatorname{sn}(u_a) \operatorname{dc}(u_a) - \left( 2-k^2 \right) u_a - 2 \sqrt{1-k^2} \operatorname{sc}(u_a) \right), \\
&\text{with } u_a = \sqrt{\frac{A}{2}}\frac{a-1}{k} = \frac{\sqrt{I_\lambda} + \sqrt{J_\lambda}}{\sqrt{2}} (a-1).
\eeq
Then we compute the second integral
\beq
\int_{a}^{1} e^{v(x)} dx &= \int_{a}^{1} \frac{dx}{\left( \frac{1}{k} \operatorname{dc}\left( \sqrt{\frac{A}{2}}\frac{x-1}{k} \right) + \frac{\sqrt{1 - k^2}}{k} \operatorname{nc}\left( \sqrt{\frac{A}{2}}\frac{x-1}{k} \right) \right)^2} \\
&= k^3 \sqrt{\frac{2}{A}} \int_{u_a}^{0} \frac{du}{\left(\operatorname{dc}(u) + \sqrt{1-k^2} \operatorname{nc}(u) \right)^2} \\
&= k^3 \sqrt{\frac{2}{A}} \int_{u_a}^{0} \frac{\left(\operatorname{dc}(u) - \sqrt{1-k^2} \operatorname{nc}(u) \right)^2}{\left(\operatorname{dc}^2(u) - (1-k^2) \operatorname{nc}^2(u) \right)^2} du.
\eeq
Since $dc^2(u) -(1-k^2)nc^2(u) = k^2$, we finally obtain
\beq
\int_{a}^{1} e^{v(x)} dx = \frac{1}{k} \sqrt{\frac{2}{A}} \int_{u_a}^{0} \left(\operatorname{dc}(u) - \sqrt{1-k^2} \operatorname{nc}(u) \right)^2 du,
\eeq
which is very similar to the previous integral eq.\eqref{eq:integral1}. We thus compute eq.\eqref{eq:int_ilambda} similarly, leading to
\beq \label{eq:int_ilambda}
\int_{a}^{1} e^{v(x)} dx = \frac{1}{k}\sqrt{\frac{2}{A}} \left( 2 \operatorname{E}(u_a) - 2 \operatorname{sn}(u_a)\operatorname{dc}(u_a) - \left( 2-k^2 \right) u_a + 2 \sqrt{1-k^2} \operatorname{sc}(u_a) \right).
\eeq
We define for $u \in ]-K(k);K(k)[$
\beq \label{eq:fg2}
f(u) &= 2 \operatorname{E}(u) - (2-k^2)u -2 \operatorname{sn}(u)\operatorname{dc}(u), \\
g(u) &= 2\operatorname{sc}(u),
\eeq
so we can now write eq.\eqref{eq:int_ilambda} and eq.\eqref{eq:int_jlambda}
\beq
\int_{a}^{1} e^{v(x)} dx &= \frac{1}{k}\sqrt{\frac{2}{A}} \left( f(u_a) + \sqrt{1-k^2} g(u_a) \right), \\
\int_{a}^{1} e^{-v(x)} dx &= \frac{1}{k}\sqrt{\frac{2}{A}} \left( f(u_a) - \sqrt{1-k^2} g(u_a) \right).
\eeq
Because $u = v + D$ we have
\beq
&\int_{a}^{1} e^v = e^{-D} \int_{a}^{1} e^u = \sqrt{\frac{I_\lambda}{J_\lambda}} \frac{\lambda_d n^-}{I_\lambda} = \frac{\lambda_d n^-}{\sqrt{I_\lambda J_\lambda}}, \\
&\int_{a}^{1} e^{-v} = e^{D} \int_{a}^{1} e^{-u} = \sqrt{\frac{J_\lambda}{I_\lambda}} \frac{\lambda_d n^+}{J_\lambda} = \frac{\lambda_d n^+}{\sqrt{I_\lambda J_\lambda}},
\eeq
and from eq.\eqref{eq:ac} and eq.\eqref{eq:k} we obtain
\beq
  \frac{1}{k}\sqrt{\frac{2}{A}} = \frac{\sqrt{I_\lambda} + \sqrt{J_\lambda}}{2\sqrt{2 I_\lambda J_\lambda}}
  \quad \text{and} \quad
  \sqrt{1-k^2} = \frac{ \sqrt{J_\lambda} - \sqrt{I_\lambda} }{\sqrt{I_\lambda} + \sqrt{J_\lambda}},
\eeq
which finally gives the system
\begin{subequations}
  \label{eq:SIJ_appendix}
  \begin{align}
    \label{eq:SI_appendix}
    \lambda_d n^- &= \frac{\sqrt{I_\lambda} + \sqrt{J_\lambda}}{2\sqrt{2}} \left( f(u_a) + \frac{ \sqrt{J_\lambda} - \sqrt{I_\lambda} }{\sqrt{I_\lambda} + \sqrt{J_\lambda}} g(u_a) \right),
    \\
    \label{eq:SJ_appendix}
    \lambda_d n^+ &= \frac{\sqrt{I_\lambda} + \sqrt{J_\lambda}}{2\sqrt{2}} \left( f(u_a) - \frac{ \sqrt{J_\lambda} - \sqrt{I_\lambda} }{\sqrt{I_\lambda} + \sqrt{J_\lambda}} g(u_a) \right).
  \end{align}
\end{subequations}
We can also notice that
\beq \label{eq:SN2}
\lambda_d N = - \frac{ \sqrt{J_\lambda} - \sqrt{I_\lambda} }{\sqrt{2}} g(u_a).
\eeq
%%%%%%%%%%%%%%%%%%%%%%%%%%%%%%%%%%%%%%%
\subsection{Appendix: Computing the leading order term $u_0$ in dimension 2} \label{appendix4}
%%%%%%%%%%%%%%%%%%%%%%%%%%%%%%%%%%%%%%%
The first term $u_0$ is the solution of
\beq\label{eq:pnpdim2_delta0}
u_0''(r) + \frac1r u_0'(r) &=& - J_\lambda e^{-u_0(r)},\\
u_0(1) = u_0'(1) &=& 0,
\eeq
which we obtained by setting $\delta = 0$ in eq.\eqref{eq:pnpdim2}. Using the change of variables
\beq
r &= e^{-t}\, \tilde{u}(t) &= u(r) + 2t,
\eeq
eq.\eqref{eq:pnpdim2} reduces to
\beq
\tilde{u}''(t) = - J_\lambda e^{-\tilde{u}(t)},
\eeq
with boundary conditions
\beq
\tilde{u}(0) = 0, \quad \tilde{u}'(0) = 2.
\eeq
We resolve here
\beq  \label{eq:dim2_int2}
\frac12 \tilde{u}'^2 = J_\lambda e^{-\tilde{u}(t)} + 2 - J_\lambda.
 \eeq
in the three cases $J_\lambda < 2$, $J_\lambda = 2$ and $J_\lambda > 2$.
%%%%%%%%%%%%%%%%%%%%%%%%%%%%%%%%%%%%%%%
\subsubsection*{Case $J_\lambda < 2$}
%%%%%%%%%%%%%%%%%%%%%%%%%%%%%%%%%%%%%%%
We integrate
\beq\label{eq:dim2_jleq2_start}
I(\tilde{u}) = \int^{\displaystyle \tilde{u}} \frac{dx}{\sqrt{J_\lambda e^{-x} + 2 -J_\lambda}}
%&= \frac{2}{\sqrt{J_\lambda}} \bigints^{\exp\left(\frac{\tilde{u}}{2}\right)} \frac{ds}{\sqrt{1+\frac{2-J_\lambda}{J_\lambda}s^2}}\\
= \frac{2}{\sqrt{2-J_\lambda}} \int^{\sqrt{\frac{2-J_\lambda}{J_\lambda}} \exp(\frac{\tilde{u}}{2})} \frac{dv}{\sqrt{1+v^2}}.
\eeq
leading to
\beq
\sqrt{\frac{2}{2-J_\lambda}} \operatorname{arsinh} \left( \sqrt{\frac{2-J_\lambda}{J_\lambda}} \exp\left( \frac{\tilde{u}}{2} \right) \right) = t + C,
\eeq
where $C = \sqrt{\frac{2}{2-J_\lambda}} \operatorname{arsinh}\left( \sqrt{\frac{2-J_\lambda}{J_\lambda}} \right)$. This leads to the simplified relation
\beq \label{eq:udim2_jleq2}
u(r) = 2 \ln\left( \frac12 \left( 1 + \frac1p \right) r^{1-p} -\frac12 \left( \frac1p - 1 \right)r^{1+p} \right),
\eeq
where $p=\sqrt{\frac{2-J_\lambda}{2}}$. To evaluate how $J_\lambda$  depends on $\lambda_d n^+$, we compute the integral in eq.\eqref{eq:defIJ} :
\beqq
\int_a^1 e^{-u(r)} r dr &=\ds \bigintsss_a^1 \frac{r dr}{\left( \frac12 \left( 1 + \frac1p \right) r^{1-p} -\frac12 \left( \frac1p - 1 \right)r^{1+p} \right)^2}\\ &= \ds4p^2 \int_a^1 \frac{r^{2p-1} dr}{\left( p+1 - (1-p)r^{2p} \right)^2} %&= 4p^2 \left[ \frac{1}{2p(1-p)\left(p+1 - (1-p)r^{2p}\right)} \right]_{r=a}^{r=1} \\
    = \ds \frac{1 - a^{2p}}{\left( 1 + a^{2p} \right)p + 1 - a^{2p}}
\eeqq
and
\beq
\lambda_d n^+ = \frac{J_\lambda \left( 1 - a^{2p} \right)}{\left( 1 + a^{2p} \right)p + 1 - a^{2p}}.
\eeq
In the limit $p \longrightarrow 0$, expanding $a^{2p}$ leads to
\beq
\lambda_d n^+ \longrightarrow \frac{2 \ln(a)}{\ln(a) - 1} \quad \text{when} \quad J_\lambda \longrightarrow 2.
\eeq
%%%%%%%%%%%%%%%%%%%%%%%%%%%%%%%%%%%%%%%%%%%%%%%%%%%555
\subsubsection*{Case $J_\lambda = 2$}
%%%%%%%%%%%%%%%%%%%%%%%%%%%%%%%%%%%%%%%%%%%%%%%%%%%555
When $J_\lambda = 2$, eq.\eqref{eq:dim2_int} becomes
\beq
\frac12 \tilde{u}'^2 = 2 e^{-\tilde{u}},
\eeq
thus
\beq
e^{\frac{\tilde{u}}{2}} \tilde{u}' = 2,
\eeq
gives the solution
\beq
\tilde{u}(t) = 2 \ln(1+t).
\eeq
Since $u(r) = \tilde{u}(-\ln(r)) + 2\ln(r)$, we obtain the solution
\beq \label{eq:dim2_jeq2_solution}
u(r) = 2 \ln(r(1-\ln(r))).
\eeq
We  evaluate $\lambda_d n^+$ by computing the integral in eq.\eqref{eq:defIJ} :
\beq
\int_a^1 e^{-u(r)} r dr = \bigintsss_a^1 \frac{dr}{r\left( 1-\ln(r) \right)^2}
%&= \left[ \frac{1}{1-\ln(r)} \right]_{r=a}^{r=1}
     = \frac{\ln(a)}{\ln(a) - 1},
\eeq
and get
\beq
\lambda_d n^+ = \frac{2 \ln(a)}{\ln(a) - 1}.
\eeq
%%%%%%%%%%%%%%%%%%%%%%%%%%%%%%%%%%%%%%%
\subsubsection*{Case  $J_\lambda > 2$}
%%%%%%%%%%%%%%%%%%%%%%%%%%%%%%%%%%%%%%%
Following \ref{eq:dim2_jleq2_start}, a direct integration leads to
\beq \label{eq:dim2_jleq2_start2}
I(\tilde{u}) = \int^{\displaystyle \tilde{u}} \frac{dx}{\sqrt{J_\lambda e^{-x} -(J_\lambda - 2)}} %\\
%&= \frac{2}{\sqrt{J_\lambda}} \bigints^{\exp\left(\frac{\tilde{u}}{2}\right)} \frac{ds}{\sqrt{1-\frac{J_\lambda-2}{J_\lambda}s^2}} \\
%&
= \frac{2}{\sqrt{J_\lambda-2}} \int^{\sqrt{\frac{J_\lambda-2}{J_\lambda}} \exp(\frac{\tilde{u}}{2})} \frac{dv}{\sqrt{1-v^2}}.
\eeq
Thus,
\beq
\sqrt{\frac{2}{J_\lambda-2}} \operatorname{arcsin} \left( \sqrt{\frac{J_\lambda-2}{J_\lambda}} \exp\left( \frac{\tilde{u}}{2} \right) \right) = t + C,
\eeq
where $C = \sqrt{\frac{2}{J_\lambda-2}} \operatorname{arcsin}\left( \sqrt{\frac{J_\lambda-2}{J_\lambda}} \right)$, leading to
\beq \label{eq:udim2_jleq2b}
u(r) = 2 \ln\left( r \left( \frac1p \sin(-p \ln(r)) + \cos(-p \ln(r)) \right) \right),
\eeq
where $p = \sqrt{\frac{J_\lambda-2}{2}}$. We can now evaluate the  relation with  $\lambda_d n^+$ in $J_\lambda$, we compute the integral in eq.\eqref{eq:defIJ} :
\beq
\int_a^1 e^{-u(r)} r dr = \int_a^1 \frac{dr}{r\left( \frac1p \sin(-p \ln(r)) + \cos(-p \ln(r)) \right)^2}
 %\\ &= \bigintsss_{\ln(a)}^0 \frac{du}{\left( \cos(pu) - \frac1p \sin(pu) \right)^2}\\
 %   &= \bigintsss_{\ln(a)}^0 \frac{du}{\cos^2(pu)\left( 1 - \frac1p \tan(pu) \right)^2}\\
    %= \left[ \frac{1}{1 - \frac1p \tan(pu)} \right]_{u = \ln(a)}^{u=0}\\&
    = \frac{1}{1 - p \cot(p \ln(a))}
\eeq
and
\beq
\lambda_d n^+ = \frac{J_\lambda}{1 - p \cot(p \ln(a))}.
\eeq
Since $J_\lambda \longrightarrow 2$, $p \longrightarrow 0$ and we obtain
\beq
\lambda_d n^+ \longrightarrow \frac{2 \ln(a)}{\ln(a) - 1} \quad \text{when} \quad J_\lambda \longrightarrow 2.
\eeq
In addition,
\beq \label{eq:dim2_jlim}
\lambda_d n^+ \longrightarrow \infty \quad \text{when} \quad J_\lambda \longrightarrow J_{lim}(a),
\eeq
where $J_{lim}(a)$ is the first positive solution of the equation
\beq
\sqrt{\frac{J-2}{2}} \cot\left( \sqrt{\frac{J-2}{2}}\ln(a) \right) = 1.
\eeq

%%%%%%%%%%%%%%%%%%%%%%%%%%%%%%%%%%%%%%%
\subsection{Appendix: Computing the first term $u_1$ of the regular perturbation }\label{sec:ap5}
%\addcontentsline{toc}{section}{\nameref{sec:ap5}}
The second term $u_1$ of the regular perturbation is the solution of
\beq \label{eq:dim2_u1p}
u_1''(r) + \frac1r u_1'(r) = e^{u_0(r)} + J_\lambda e^{-u_0(r)} u_1(r),
\eeq
with  boundary conditions
\beq\label{eq:dim2_u1_initial}
u_1(1) = u_1'(1) = 0.
\eeq
We distinguish three cases $J_\lambda < 2$, $J_\lambda = 2$ and $J_\lambda > 2$. For $J_\lambda < 2$, the homogeneous equation is
\beq \label{eq:dim2_jleq2_homo}
u''(r) + \frac1r u'(r) -\frac{8 p^2 (1-p^2)}{r^2\left( (p+1) r^{-p} - (1-p) r^p \right)^2} u(r) = 0,
\eeq
where $p = \sqrt{1-\frac{J_\lambda}{2}}$. We use the change of variable $x = r^p$ and $u(r) = v(x)$, to transform the equation into
\beq
v''(x) + \frac1r v'(x) &- \frac{8 q}{(q - x^2)^2} v(r) = 0,\\
v(1) &= v'(1) = 0,
\eeq
where $q = \frac{1+p}{1-p}$. The two independent solutions  are
\beq
y_1(x) &=& \frac{x^2 + q}{x^2 - q},\\
y_2(x) &=& y_1(x) \ln(x) - 1,
\eeq
thus the solutions to eq.\eqref{eq:dim2_jleq2_homo} are
\beq
Y_1(r) &=& \frac{r^{2p} + q}{r^{2p} - q},\\
Y_2(r) &=& p Y_1(r) \ln(r) - 1.
\eeq
Finally, the general solution of eq.\eqref{eq:dim2_u1p} with initial conditions \eqref{eq:dim2_u1_initial} is
\beq
u_1(r) = (\lambda(r) + A) Y_1(r) + (\mu(r) + B) Y_2(r),
\eeq
where
\beq
\mu(r) &=& \frac{(1-p)^2}{4 p^3 (4+2p)} r^{4+2p} - \frac{(1+p)^2}{4 p^3 (4-2p)} r^{4-2p},\\
\lambda(r)&=& - p \mu(r) \ln(r) - \frac{1-p^2}{8 p^3} r^4 + \frac{(1-p)^2 (4+3p)}{4 p^3 (4+2p)^2} r^{4+2p} + \frac{(1+p)^2 (4-3p)}{4 p^3 (4-2p)^2} r^{4-2p},\nn\\
A &=& - \frac18 \frac{p^4 - 23 p^2 + 40}{p (p^2 - 4)^2},\, B = -\frac14 \frac{p^2 + 5}{p^2 (p^2 - 4)}.\nn
\eeq
When $J_\lambda = 2$, eq.\eqref{eq:dim2_u1} becomes
\beq
u_1''(r) + \frac1r u_1'(r) - \frac{2}{r^2 (1-\ln(r))^2} u_1(r) = r^2 (1-\ln(r))^2.
\eeq
The solution is$ u_1(r) = \left( A + \lambda(r) \right) \left( 1 - \ln(r) \right)^2 + \frac{B + \mu(r)}{1-\ln(r)},$
where
\beqq
A &=& \frac{5}{48}, \,B  = -\frac{103}{384} \\
\lambda(r) &=& \frac{r^4}{12} \left( -\frac{5}{4} + \ln(r) \right)\\
\mu(r) &=& \frac{r^4}{384} \left( 32 \ln(r)^4 - 160 \ln(r)^3 + 312 \ln(r)^2 - 284 \ln(r) + 103 \right).
\eeqq
Finally, when $J_\lambda > 2$, the homogeneous eq.\eqref{eq:dim2_u1} is
\beq  \label{eq:dim2_u1_jgeq2_homo}
u_1''(r) + \frac1r u_1'(r) - \frac{2 (1+p^2)}{r^2 \cos^2(p\ln(r))\left( 1-\frac1p \tan(p\ln(r)) \right)^2} u_1(r) = 0,
\eeq
where $p = \sqrt{\frac{J_\lambda}{2} - 1}$. We use the change of variable $x = \tan(p \ln(r))$ and $v(x) = u(r)$ to get
\beq
v''(x) + \frac{2 x}{1+x^2} v'(x) - \frac{2 (1+p^2)}{(p-x)^2(1+x^2)} v(x) = 0.
\eeq
The independent solutions are
\beqq
y_1(x) &= \frac{1+px}{x-p},\\
y_2(x) &= y_1(x) \arctan(x) - 1.
\eeqq
Therefore the solutions to the homogenous eq. \eqref{eq:dim2_u1_jgeq2_homo} are
\beq
Y_1(r) &= \ds \frac{1 + p\tan(p\ln(r))}{\tan(p\ln(r))-p},\\
Y_2(r) &= p Y_1(r) \ln(r) - 1.
\eeq
Finally, the solution of eq.\eqref{eq:dim2_u1} with initial conditions \eqref{eq:dim2_u1_initial} is
\beq
u_1(r) = (\lambda(r) + A) Y_1(r) + (\mu(r) + B) Y_2(r),
\eeq
where
\beqq
\mu(r) &=& \frac{r^4}{p^3 (16+4p^2)}\left( \left(5p - p^3\right) \cos(2 p \ln(r)) - (2 - 4p^2) \sin(2 p \ln(r))\right),\\
\lambda(r) &=& -p \mu(r) \ln(r) - \frac{(1+p^2)r^4}{8 p^3} - \frac{7 p^4 + 8 p^2 - 8}{4 p^3 (p^2 + 4)^2} r^4 \cos(2 p \ln(r))\\ &-& \frac{3p^4 -15p^2-36}{8p^2 (p^2 + 4)^2} r^4 \sin(2 p \ln(r)),\\
A &=& \frac18 \frac{p^4 + 23p^2 + 40}{p (p^2+4)^2}, \\
B &=& \frac14 \frac{p^2 - 5}{p^2 (p^2 + 4)}.
\eeqq
%%%%%%%%%%%%%%%%%%%%%%%%%%%%%%%%%%%%%%%%%%%%%%%%%%%%%%%%%%%%%%%%%%%%%%%
\newpage
\normalem
%%%%%%%%%%%%%%%%%%%%%%%%%%%%%%%%%%%
\bibliographystyle{ieeetr}%{plain}%\bibliographystyle{ieeetr}%{apsrev4-1} %{ieeetr}
\bibliography{Cusp_3Ddh,RMPbiblio4newN,RMPbiblio4newN2}
\end{document}